\documentclass[aps,prd,twocolumn,floatfix,nofootinbib,superscriptaddress]{revtex4-1}
\usepackage[colorlinks,linkcolor=blue,anchorcolor=blue,citecolor=blue,urlcolor=blue,breaklinks=true]{hyperref}
\usepackage[libertine]{newtxmath}
\usepackage{graphicx}
\usepackage{slashed}
\usepackage{amsfonts}
\usepackage{amsmath}
\usepackage{blindtext}
\usepackage{microtype}

\newcommand{\vect}[1]{\boldsymbol{#1}}
\newcommand{\ph}[1]{\phantom{#1}}
\newcommand{\sh}[1]{\slashed{#1}}

\graphicspath{{.}{figures/}} 

\def\emdash{\hs{1pt}---\hs{1pt}}

\def\hs{\hspace}

\def\no{\nonumber}

\def\lf{\left}
\def\rg{\right}

\begin{document}

\title{Spatial and Momentum Imaging of the Pion and Kaon}

\author{Chao Shi}
\email[]{cshi@nuaa.edu.cn}
\affiliation{Department of Nuclear Science and Technology, Nanjing University of Aeronautics and Astronautics, Nanjing 210016, China}

\affiliation{Collaborative Innovation Center of Radiation Medicine of Jiangsu Higher Education Institutions,
Nanjing 211106, China}

\author{Kyle Bednar}
\affiliation{Center for Nuclear Research, Department of Physics, Kent State University, Kent OH 44242 USA}

\author{Ian C. Clo\"et}
\affiliation{Physics Division, Argonne National Laboratory, Argonne, IL 60439 USA}

\author{Adam Freese}
\affiliation{Physics Division, Argonne National Laboratory, Argonne, IL 60439 USA}

\begin{abstract}
We determine the leading Fock state light front wave functions (LFWFs) of the pion and kaon via light front projections of the  covariant Bethe-Salpeter wave function. Using these LFWFs we study the multi-dimensional images of the valence quarks in the pion and kaon that are provided by their generalized parton distribution functions (GPDs) and transverse momentum dependent parton distribution functions (TMDs). Moments of the GPDs are taken to obtain the electromagnetic and gravitational form factors of the pion and kaon, and comparisons to available experimental and lattice data are made. Highlights from this study include predictions that the mean-squared impact parameter for the quarks in the pion and kaon are: $\langle \vect{b}_T^2\rangle_{u}^\pi=0.11$\,fm$^2$, $\langle  \vect{b}_T^2\rangle_{s}^K=0.08$\,fm$^2$, and $\langle \vect{b}_T^2\rangle_{u}^K=0.13$\,fm$^2$, and therefore the $s$ quark in the kaon is much closer to the center of transverse momentum than the $u$ quark. From the electromagnetic and gravitational form factors we find that the light-cone energy radii are about 60\% smaller than the light-cone charge radii for each quark sector in the pion and kaon. A quantitative measure of the importance of the leading Fock state is obtained via comparison with a full DSE calculation (containing an infinite tower of Fock states) for the pion form factor.
\end{abstract}
\maketitle

\section{INTRODUCTION\label{intro}}
Multi-dimensional images of the partonic structure of hadrons are provided by the generalized parton distribution functions (GPDs)~\cite{Mueller:1998fv,Ji:1996nm,Radyushkin:1997ki} and transverse momentum dependent parton distributions functions (TMDs)~\cite{Collins:2003fm}. These images encode abundant structural information about hadrons, e.g., the GPDs provide a unified description of form factors and parton distribution functions (PDFs), where the former is related to a hadron's spatial extent and the latter describes the light-cone momentum distribution of partons~\cite{Burkardt:2000za,Burkardt:2002hr} within a hadron. Through $x$-weighted moments the GPDs are connected with hadron matrix elements of the energy-momentum tensor, and therefore shed-light on the spin, energy, and pressure distributions within hadrons~\cite{Ji:1996nm,Burkert:2018bqq}. Experimentally, GPDs are accessible through hard exclusive processes like deeply virtual Compton scattering (DVCS) or deeply virtual meson production (DVMP). TMDs illustrate the transverse motion of the partons in 3-dimensional momentum space, and therefore complement GPDs. Hadron TMDs can be extracted from semi-inclusive deep inelastic scattering (SIDIS) or Drell-Yan processes.

Calculating GPDs and TMDs directly from the fundamental theory, quantum chromodynamics (QCD), has proven very challenging. Lattice QCD has typically been limited to certain aspects of GPDs and TMDs, such as low $x$-weighted moments, together with their $\vect{k}_T^2$-dependence for TMDs or $t$-dependence~\cite{Musch:2010ka, Musch:2011er,Yoon:2017qzo} for GPDs~\cite{Gockeler:2003jfa,Hagler:2007xi,Alexandrou:2019ali}. However, new approaches, such as Large-Momentum Effective Theory (LaMET)~\cite{Ji:2014hxa,Ji:2015qla,Ji:2018hvs,Ebert:2019okf}, now  enable lattice QCD to reveal much richer information on GPDs and TMDs. Model calculations are also crucial, as they can help provide an intuitive picture of the GPDs and TMDs. For instance, using the Nambu--Jona-Lasinio model or the spectral quark model one can calculate the full pion GPD over the entire kinematic range $|x|<1$, $|\xi|<1$~\cite{Broniowski:2007si}. Such a calculation is based on non-perturbative covariant Feynman diagrams. An alternative approach is the light front QCD framework, where the GPDs and TMDs are determined through overlap representations in terms of light front wave functions (LFWFs)~\cite{Diehl:2000xz,Diehl:2003ny}. The unknown elements are then the non-perturbative LFWFs of hadrons. In this work, we will determine the LFWFs of the pion and kaon from a beyond rainbow-ladder Dyson-Schwinger equations (DSE) calculation, and then study the GPDs and TMDs obtained from these LFWFs using overlap representations. The pion and kaon are of particular interest as they emerge as the Goldstone bosons associated with dynamical chiral symmetry breaking (DCSB) in QCD. Pion and kaon GPDs and TMDs are also experimentally accessible\emdash in principle\emdash via hadron-hadron collisions using pion and kaon beams, or through interactions with the virtual meson cloud around nucleon targets~\cite{Aghasyan:2017jop,pieic:2018}. 

To calculate the LFWFs the standard approach is to diagonalize the light-cone QCD Hamiltonian within the light-cone QCD formalism~\cite{Brodsky:1997de}. However, in practice this is numerically very difficult for exact QCD in four spacetime dimensions, and therefore effective interactions are usually adopted~\cite{Brodsky:2014yha,Gutsche:2014zua,Jia:2018ary,Lan:2019vui,Lan:2019rba,Li:2017mlw}. An alternative approach is to solve the covariant Bethe-Salpeter equation and project the Bethe-Salpeter wave functions onto the light front. This idea originates from a model calculation by 't Hooft~\cite{tHooft:1974pnl} and was also used by authors in Refs.~\cite{Liu:1992dg,Heinzl:2000ht}. In a recent work the pion's leading Fock state LFWFs were obtained from its Bethe-Salpeter wave function provided by a DSE calculation~\cite{Shi:2018zqd}, and used to study the pion TMD. In this paper we extend this work to the kaon LFWFs and TMDs, and also the study of pion and kaon GPDs.

In the past few decades the DSE framework has been applied extensively to hadron physics~\cite{Roberts:1994dr,Cloet:2013jya}.  By solving the quark's gap equation and meson's Bethe-Salpeter equation, one obtains the covariant Bethe-Salpeter wave function, from which hadron properties can be determined. The DSEs respect the (approximate) chiral symmetry in the light quark sector, and its dynamical breaking, as demonstrated by satisfying the axial-vector Ward-Takahashi identity (AV-WTI)~\cite{Maris:1997hd}. Therefore, the extracted LFWFs encode the effects from DCSB and provide a realistic description of the Goldstone bosons at leading-order in the Fock state expansion. Therefore, DSE predictions for the two-particle LFWFs of pion and kaon can provide important insights into the structure of QCD's Goldstone bosons.

This paper is organized as follows: In Sec.~\ref{sec:LFWF} we determine the LFWFs of pion and kaon from their Bethe-Salpeter wave functions. We then study their GPDs and related form factors in Sec.~\ref{sec:GPD} and Sec.~\ref{sec:FF}. The unpolarized TMDs are determined in Sec.~\ref{sec:TMD} and a conclusion is given in Sec.~\ref{sec:con}.

\section{PION AND KAON LIGHT FRONT WAVE FUNCTIONS\label{sec:LFWF}}
In light-front QCD hadron states are generally described by a tower of Fock states in a Fock state expansion~\cite{Brodsky:1997de,Heinzl:2000ht}. For a meson with valence quark $f$ and valence  anti-quark $\bar{h}$ the minimal (2-particle) Fock-state configuration is given by~\cite{Li:2017mlw,Jia:2018ary}
\begin{align}\label{eq:LFWF1}
|M\rangle &= \sum_{\lambda_1,\lambda_2}\int \frac{d^2 \vect{k}_T}{(2\pi)^3}\,\frac{dx}{2\sqrt{x\bar{x}}}\, \frac{\delta_{ij}}{\sqrt{3}} \nonumber \\
&\hspace{10mm} \Phi_{\lambda_1,\lambda_2}(x,\vect{k}_T)\, b^\dagger_{f,\lambda_1,i}(x,\vect{k}_T)\, d_{h,\lambda_2,j}^\dagger(\bar{x},\bar{\vect{k}}_T)|0\rangle.
\end{align}
where $\vect{k}_T$ is the transverse momentum of the quark $f$ [in a frame where the meson's transverse momentum vanishes ($\vect{P}_T = 0$)], $\bar{\vect{k}}_T=-\vect{k}_T$, $x=\frac{k^+}{P^+}$ is the light-cone momentum fraction of the active quark, and $\bar{x}=1-x$. The quark helicity is labelled by $\lambda_i = (\uparrow,\downarrow)$ and $\delta_{ij}/\sqrt{3}$ is a color factor. 

Ref.~\cite{Burkardt:2002uc} showed that for pseudo-scalar mesons there are two independent light front wave functions for the leading Fock state, labeled by $\psi_0(x,\vect{k}_T^2)$ with $l_z=0$ and $\psi_1(x,\vect{k}_T^2)$ with $|l_z|=1$. The 2-particle Fock-state configuration is then given by
\begin{align}
\left|M \right> &=\left|M\right>_{l_z=0}+\left|M\right>_{|l_z|=1},\label{eq:LFWF2-0}
\end{align}
where
\begin{align}
\label{eq:LFWF2-1}
&|M\rangle_{l_z=0} = i\int \frac{d^2 \vect{k}_T}{2(2 \pi)^3}\frac{dx}{\sqrt{x\bar{x}}}\ \psi_0(x,\vect{k}_T^2)\ \frac{\delta_{ij}}{\sqrt{3}} \frac{1}{\sqrt{2}}\nonumber\\
&\hspace{10mm}
[b^\dagger_{f \uparrow i}(x,\vect{k}_T)d^\dagger_{h \downarrow j}(\bar{x},\bar{\vect{k}}_T)-b^\dagger_{f \downarrow i}(x,\vect{k}_T)d^\dagger_{h \uparrow j}(\bar{x},\bar{\vect{k}}_T)]|0\rangle,  \\
&|M\rangle_{|l_z|=1} = i\int \frac{d^2 \vect{k}_T}{2(2 \pi)^3}\frac{dx}{\sqrt{x\bar{x}}}\ \psi_1(x,\vect{k}_T^2)\ \frac{\delta_{ij}}{\sqrt{3}} \frac{1}{\sqrt{2}}
\nonumber\\
&\hspace{3mm}
[k_T^-\,b^\dagger_{f \uparrow i}(x,\vect{k}_T)d^\dagger_{h \uparrow j}(\bar{x},\bar{\vect{k}}_T)
+ k_T^+\,b^\dagger_{f \downarrow i}(x,\vect{k}_T)d^\dagger_{h \downarrow j}(\bar{x},\bar{\vect{k}}_T)]|0\rangle, \label{eq:LFWF2-2}
\end{align}
where $k_T^{\pm}=k^1\pm ik^2$. The LFWFs are obtained from the Bethe-Salpeter wave function via the light front projections~\cite{Mezrag:2016hnp,Shi:2018zqd}
\begin{align}
\label{eq:psi0}
\psi_0(x,\vect{k}_T^2)&= \ph{-}\sqrt{3}\,i\!\int \frac{dk^+dk^-}{2\,\pi} \nonumber \\
& \hspace{11mm} 
\textrm{Tr}_D\!\left[ \gamma^+ \gamma_5 \chi(k,P)\right]  \delta\left(x\,P^+ -k^+\right),  \\
\label{eq:psi1}
\psi_1(x,\vect{k}_T^2)&= -\sqrt{3}\,i\!\int \frac{dk^+dk^-}{2\,\pi}\,  \frac{1}{\vect{k}_T^2} \nonumber \\ 
& \hspace{11mm} 
\textrm{Tr}_D\left[ i\sigma_{+ i}\, k_{T}^i\, \gamma_5\, \chi(k,P) \right] 
 \delta\left(x\,P^+ -k^+\right),   
\end{align}
where the trace is over Dirac indices. The Bethe-Salpeter wave function is defined by $\chi_{f\bar{h}}(k,P) = \int d^4 z\ e^{-ik \cdot z}\, \langle 0|\mathcal{T} f(z)\,\bar{h}(0)| M(P)\rangle$~\cite{Itzykson:1980rh,Gromes:1992ph}, and can be expressed as $\chi_{f\bar{h}}(k,P) = S_f(k+P/2)\,\Gamma_{f\bar{h}}(k,P)\,S_h(k-P/2)$, where $S(k)$ is the dressed quark propagator and $\Gamma(k,P)$ the meson's Bethe-Salpeter amplitude~\cite{LlewellynSmith:1969az,Roberts:1994dr}. 

In the framework of the DSEs  $S(k)$ and $\Gamma(k,P)$ are obtained by solving the quark gap equation and Bethe-Salpeter equation, respectively. For non-singlet pseudo-scalar mesons the AV-WTI should be preserved by carefully selecting truncation schemes. The simplest symmetry-preserving DSE truncation is rainbow-ladder (RL) and has achieved many successes in the study of hadron properties~\cite{Maris:1997tm,Maris:1999nt,Maris:2000sk}. A modern extension known as the DCSB-improved (DB) truncation improves upon the RL truncation and provides more realistic description of the pion, kaon, and other hadrons~\cite{Cloet:2013jya}. In this work we employ the existing DB-kernel solution parameterized in Refs.~\cite{Chang:2013pq, Shi:2014uwa, Shi:2015esa}. Further details about this DSE truncation are given in App.~\ref{sec:para}.

To obtain the pion and kaon LFWFs we first determine an arbitrary $\vect{k}_T^2$-dependent moment defined by
\begin{align}
\left< x^m \right>_{l_z}(\vect{k}^2_T) &= \int_0^1 dx\, x^m\, \psi_{l_z}(x,\vect{k}_T^2).\label{eq:momsM}
\end{align}
These can be directly calculated using Eqs.~(\ref{eq:psi0}) and (\ref{eq:psi1}), that is
\begin{align}
\langle x^m \rangle_{0}(\vect{k}_T^2)&=\ph{-}\frac{\sqrt{3}\,i\!}{|P^+|}\int \frac{dk^+dk^-}{2\,\pi} \left (\frac{k^+}{P^+}\right )^m \nonumber \\
&\hspace{25mm} \textrm{Tr}_{\textrm{D}}  [ \gamma^+ \gamma_5\, \chi(k^+,k^-;\vect{k}_T,P)],   \label{eq:mom1}\\
\langle x^m \rangle_{1}(\vect{k}_T^2)&=-\frac{\sqrt{3}\,i\!}{|P^+|\,\vect{k}_T^2}\int \frac{dk^+dk^-}{2\,\pi} \left (\frac{k^+}{P^+}\right )^m \nonumber \\
& \hspace{20mm}  \textrm{Tr} _{\textrm{D}} [i\sigma_{+ i}\,k_T^{i}\, \chi(k^+,k^-;\vect{k}_T,P)].  \label{eq:mom2}
\end{align}
Since we have an analytical form for $\chi(k,P)$ obtained by parametrizing the numerical DSE solution, the two-dimensional momentum integrations can be completed with the help of Feynman parametrization.  In practice, we transform the integration variables to rewrite the integral in the form
 \begin{align}
\label{eq:ced}
\left< x^m \right>_{l_z}(\vect{k}^2_T) = \int_0^1 d\alpha\, \alpha^m \int d\beta d\gamma\,  f_{l_z}(\alpha,\vect{k}_T^2,\beta,\gamma).
\end{align}
Comparison with Eq.~(\ref{eq:momsM}) then reveals that the LFWFs are identified as $\psi_{l_z}(x,\vect{k}_T^2)= \int d\beta d\gamma \, f_{l_z}(x,\vect{k}_T^2,\beta,\gamma)$. 

\begin{figure}[tbp]
\centering\includegraphics[width=\columnwidth]{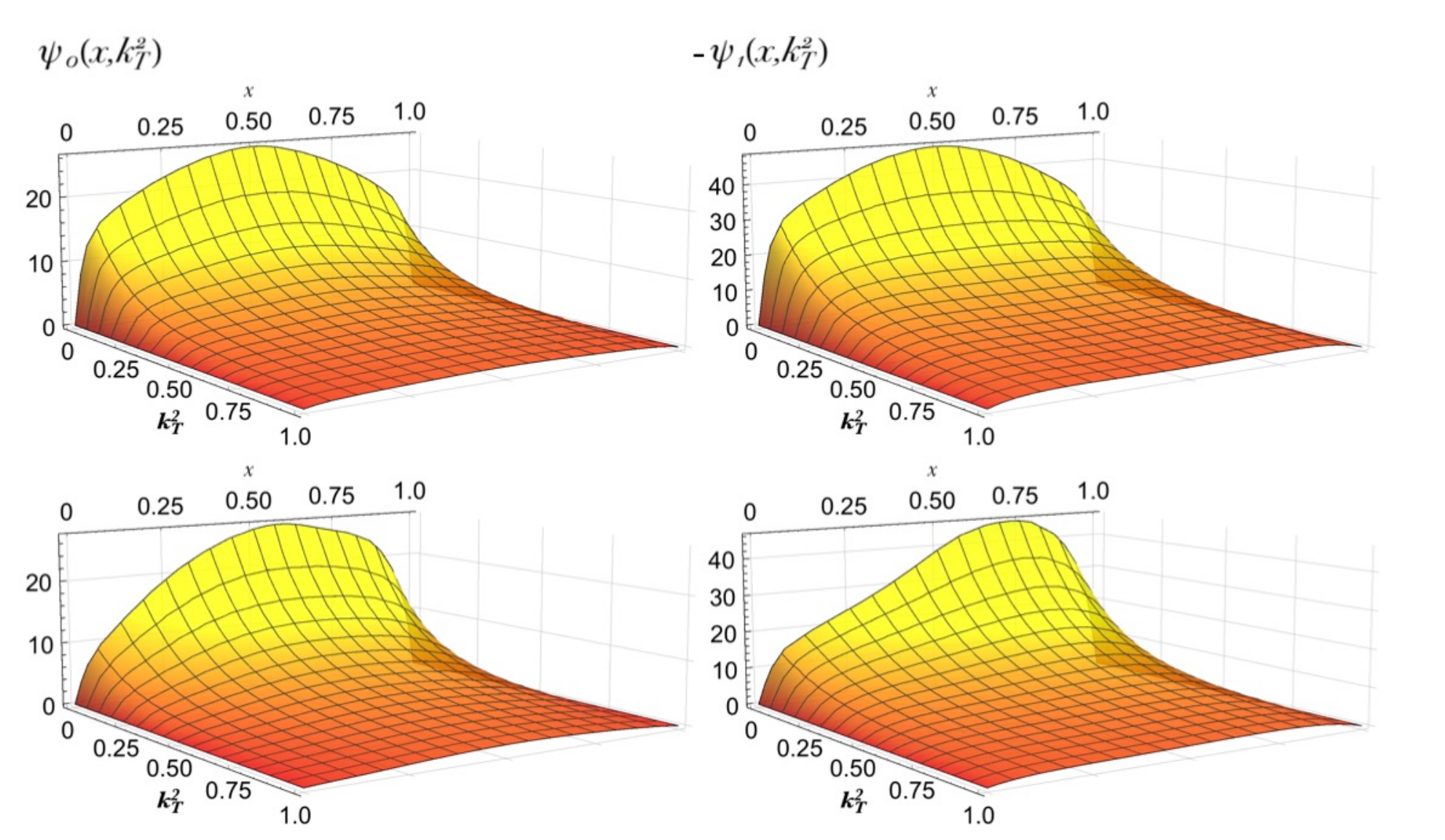}
\caption{The top row gives the LFWFs for pion and the bottom row gives the kaon results. The left column is $\psi_0(x,\vect{k}_T^2)$ and the right column is $\psi_1(x,\vect{k}_T^2)$, where  $\vect{k}^2_T$ is in GeV$^2$.
}
\label{fig:psis}
\end{figure}

\begin{figure}[tbp]
\centering\includegraphics[width=\columnwidth]{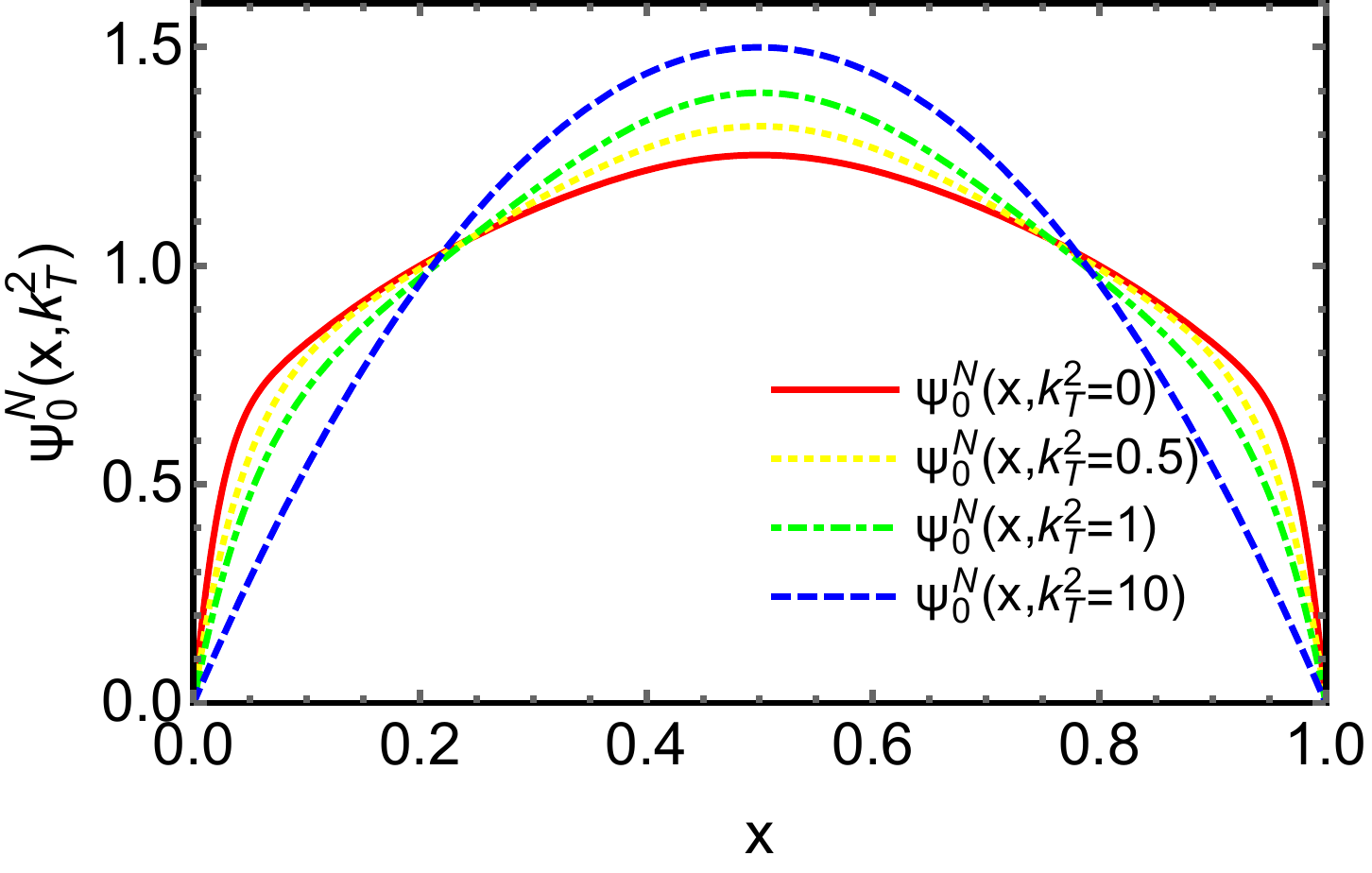}
\caption{Pion's spin-anti-parallel LFWF $\psi_0(x,\vect{k}_T^2)$ at different values of $\vect{k}_T^2$, normalized to $\psi^N_0(x,\vect{k}_T^2)=\frac{\psi_0(x,\vect{k}_T^2)}{\int_0^1 dx \psi_0(x,\vect{k}_T^2)}$.
}
\label{fig:shape}
\end{figure}

We present plots of the leading Fock state LFWFs for the pion and kaon in Fig.~\ref{fig:psis}. For concreteness, we focus our discussion to the case of $\pi^-$ and $K^-$, so the $d$ and $s$ are the valence quarks and $\bar{u}$ is valence anti-quark.  In general we find that all the LFWFs are smooth functions decaying as $\vect{k}_T^2$ increases or $x$ approaches the end-points. As expected for light mesons, the $x$-dependence of the LFWFs is broad at low $\vect{k}_T^2$ and get narrower as $\vect{k}_T^2$ increases, approaching an asymptotic form for large $\vect{k}_T^2$ proportional to $x(1-x)$. Fig.~\ref{fig:shape} provides an example of how the $x$-dependence of $\psi_0(x,\vect{k}_T^2)$ changes with $\vect{k}_T^2$. The strong support of the LFWFs at infrared $\vect{k}_T^2$ originates from the strength of the covariant Bethe-Salpeter wave functions at low $|\vect{k}_T|$, which is closely connected to DCSB, as illustrated model-independently in~Ref.~\cite{Maris:1997hd}. Therefore, our LFWFs faithfully inherit the DCSB property from the covariant DSEs calculation. At large $\vect{k}_T^2$, the LFWFs decay as $\psi_0(x,\vect{k}_T^2) \sim 1/\vect{k}_T^2$ and $\psi_1(x,\vect{k}_T^2) \sim 1/\vect{k}_T^4$, in line with the perturbative QCD expectations~\cite{Ji:2003fw}. The effects of SU(3) flavor symmetry breaking are clearly apparent in the kaon, as the heavier $s$ quark gains more support at large $x$ and the LFWFs become skewed. This indicates that the $s$ quark carries more of the kaon's light-cone momentum fraction. However, these SU(3) flavor symmetry breaking effects diminish as $\vect{k}_T^2$ increases.
Further analysis of these effects will be given in later sections when GPD and TMD results are presented.

The LFWFs are normalized so that the quark number sum rule $\int_0^1 dx\, f(x;\mu_0) = 1$ is satisfied. Therefore, with only the leading Fock state the valence quark distribution function $f(x;\mu_0)$ is given by
\begin{align}
f(x;\mu_0) = \int\frac{d^2 \vect{k}_T}{(2\pi)^3} 
\lf[\lf|\psi_0(x,\vect{k}_T^2)\rg|^2 + \vect{k}_T^2\lf|\psi_1(x,\vect{k}_T^2)\rg|^2 \rg]. 
\label{eq:fdef}
\end{align}
This approximation to the full valence quark distribution function is best at a low hadronic scale $\mu_0$, which in Ref.~\cite{Shi:2018zqd} was determined to be $\mu_0 = 520\,$MeV. In a non-relativistic system $\psi_1(x,\vect{k}_T^2)$ would vanish because the quarks are in a relative $p$-wave, however we find that the contribution to the quark number sum rule from $\psi_1(x,\vect{k}_T^2)$ equals 0.36 for the pion and 0.31 for the kaon. Therefore, we find that the valence quarks in both the pion and kaon are highly relativistic. Importantly, the relative strength between $\psi_0(x,\vect{k}_T^2)$ and $\psi_1(x,\vect{k}_T^2)$ in our approach is completely determined by the Bethe-Salpeter wave function, which itself is governed by the underlying quark-gluon interaction. The significant contribution of $\psi_1(x,\vect{k}_T^2)$ to observables likely also implies that higher Fock states may not be negligible in a more realistic calculation. Nevertheless, the higher Fock states are much more difficult to calculate and are beyond the scope of this work.

\section{GPDS AT ZERO SKEWNESS}\label{sec:GPD}
The leading twist spin-independent quark GPD for a meson $M$ is defined in light-cone gauge as
\begin{align}
H^q_M(x,\xi,t) &= \frac{1}{2}\int\frac{dz^-}{2\pi}\,e^{ixP^+z^-}  \nonumber \\
&\hs*{10mm}
\left< P+\tfrac{\Delta}{2}\left|\bar{\psi}^q(-\tfrac{z^-}{2})\,\gamma^+\,\psi^q(\tfrac{z^-}{2})\right|P-\tfrac{\Delta}{2}\right>, \label{eq:Hdef}
\end{align}
where the gauge link is unity~\cite{Ji:1996nm,Diehl:2003ny}, $x$ denotes the parton's averaged light-cone momentum fraction, the skewness parameter is $\xi=-\frac{\Delta^+}{2 P^+}$, and the momentum transfer $t = \Delta^2 = -\frac{4\xi^2 m_M^2+\vect{\Delta}_T^2}{1-\xi^2}$.  The physical support region of $H^q_M(x,\xi,t)$ is given by $x \in[-1,1]$, $\xi \in [-1,1]$ and $t<-\frac{4\xi^2 m_M^2}{1-\xi^2}$. GPDs have two distinct domains, where $|x|<|\xi|$ is the Efremov--Radyushkin--Brodsky--Lepage (ERBL) region and $1>|x|>|\xi|$ is the Dokshitzer--Gribov--Lipatov--Altarelli--Parisi (DGLAP) region, following the pattern of their evolution with scale $\mu$, which is implicit in the definition Eq.~(\ref{eq:Hdef}). 

To calculate $H^q_M(x,\xi,t)$ we employ its light front overlap representation. This result can be obtained using light-cone quantization and expanding the quark field in Eq.~(\ref{eq:Hdef}) using the canonical field mode expansion and the hadron state ket using a Fock state expansion. Contracting all the operators, one gets the light front overlap representation of the GPD in terms of LFWFs. However, in the ERBL region this requires the overlap of LFWFs with different numbers of constituents, i.e., $N$ and $N+2$. The ERBL region is therefore inaccessible is a leading Fock state expansion due to the lack of a 4-particle Fock state. 

In a meson $M$ with active quark $f$, the GPD $H^f_M(x,\xi,t)$ in the DGLAP region can be expressed as the overlap of LFWFs~\cite{Diehl:2003ny,Diehl:2000xz,Mezrag:2016hnp,Chouika:2016cmv}
\begin{align}
\label{eq:Hoverlap}
H^f_M(x,\xi,t)&=\int \frac{d^2\vect{k}_T}{(2\pi)^3}\big[ \psi_0^*(\hat{x},\hat{\vect{k}}_T)\,\psi_0(\tilde{x},\tilde{\vect{k}}_T)\nonumber \\
&\hspace{17mm}
+\hat{\vect{k}}_T \cdot \tilde{\vect{k}}_T\,\psi_1^*(\hat{x},\hat{\vect{k}}_T)\,\psi_1(\tilde{x},\tilde{\vect{k}}_T) \big],
\end{align}
with $\hat{x}=\frac{x-\xi}{1-\xi}$, $\tilde{x}=\frac{x+\xi}{1+\xi}$, $\hat{\vect{k}}_T=\vect{k}_T+\frac{1-x}{1-\xi}\frac{\vect{\Delta}_T}{2}$ and $\tilde{\vect{k}}_T=\vect{k}_T-\frac{1-x}{1+\xi}\frac{\vect{\Delta}_T}{2}$. For the active anti-quark $\bar{h}$, the GPD can be obtained analogously as~\cite{Diehl:2003ny}
\begin{align}
H^h_M(x,\xi,t)&=-\int \frac{d^2\vect{k}_T}{(2\pi)^3}\big[ \psi_0^*(\hat{x}',\hat{\vect{k}}'_T)\psi_0(\tilde{x}',\tilde{\vect{k}}'_T)\nonumber \\
&\hspace{17mm}
+\hat{\vect{k}}'_T \cdot \tilde{\vect{k}}'_T\psi_1^*(\hat{x}',\hat{\vect{k}}'_T)\psi_1(\tilde{x}',\tilde{\vect{k}}'_T) \big],
\end{align}
with $\hat{x}'=1-\frac{-x-\xi}{1-\xi}$, $\tilde{x}'=1-\frac{-x+\xi}{1+\xi}$, $\hat{\vect{k}}'_T=\vect{k}_T+\frac{1+x}{1-\xi}\frac{\vect{\Delta}_T}{2}$ and $\tilde{\vect{k}}'_T=\vect{k}_T-\frac{1+x}{1+\xi}\frac{\vect{\Delta}_T}{2}$. In the absence of an accessible ERBL region we limit our study to zero skewness, $H_M^q(x,0,t)$, which still allows access to many interesting quantities, e.g., the collinear PDF $f_M^q(x)$, the impact parameter dependent parton distributions (IPDs) $\rho_M^q(x,\vect{b}_T^2)$, the electromagnetic form factor (EMFF) $F_M(t)$, and the gravitational form factor (GFF) $A_{2,0}^{q,M}(t)$.

\begin{figure}[tbp]
\includegraphics[width=\columnwidth]{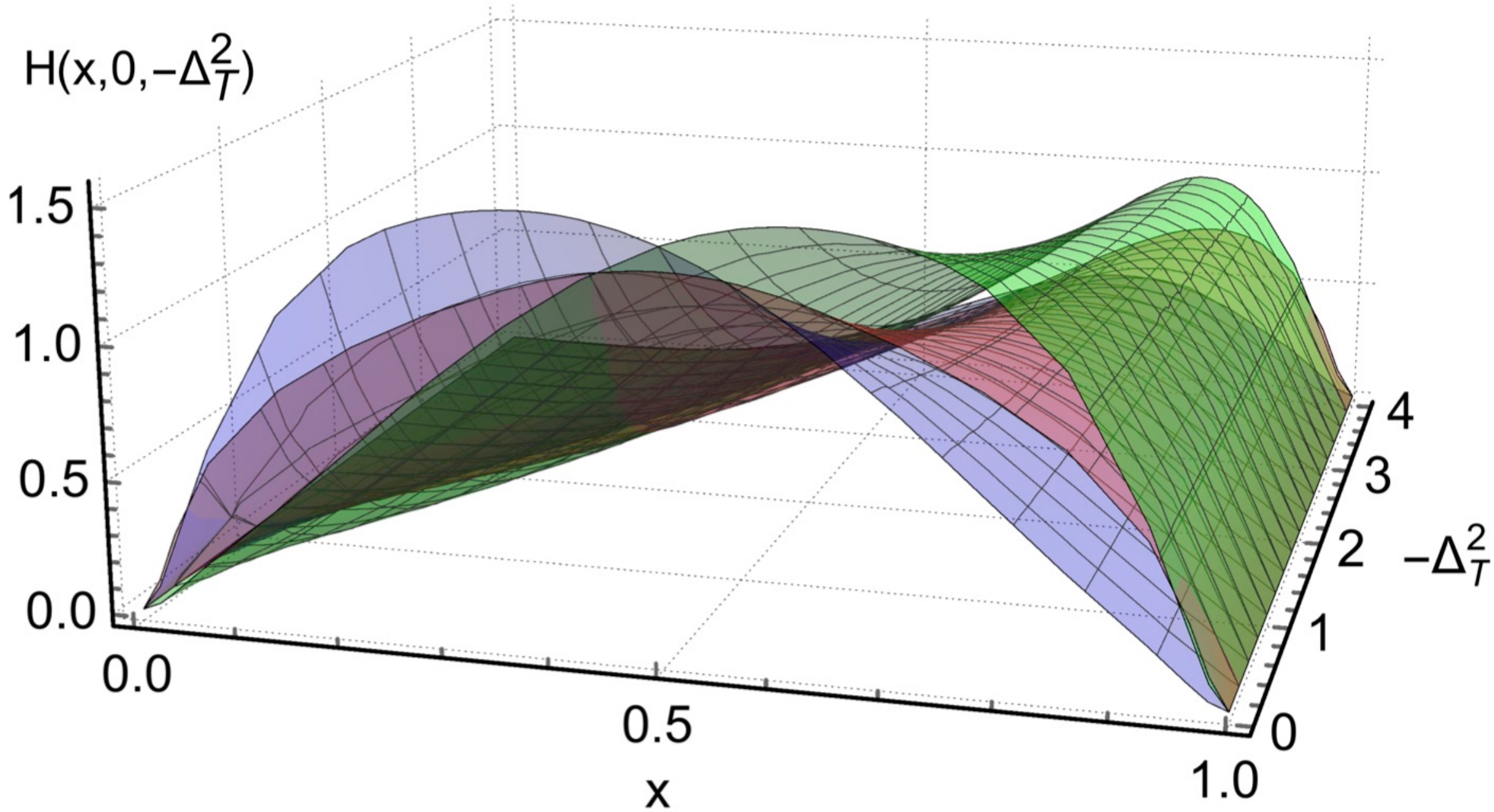} \\
\caption{$H_M^q(x,\xi=0,t)$ for pion and kaon at the model scale ($\mu_0=520$\,MeV). The green surface (upper at $x \sim 0.8$) is for the $s$ quark in kaon, blue surface (lower at $x \sim 0.8$) is for the $\bar{u}$ in kaon, and the red surface (middle at $x \sim 0.8$) is for pion.}
\label{fig:H0}
\end{figure}

We present the GPDs at the model scale $\mu_0$ in Fig.~\ref{fig:H0}. For the ease of comparison, for antiquark $\bar{h}$ we plot $-H_M^h(-x,0,t)$. Since the GPD reduces to the PDF at zero momentum transfer, i.e., $H^q(x,0,0)=f^q(x)$. The initial scale $\mu_0$ is determined as follows (see Ref.~\cite{Shi:2018zqd}): At the scale of $Q^2 = 4\,$GeV$^2$, the $\pi N$ Drell-Yan analysis gives averaged momentum fraction of valence quark distribution in pion as $2\,\langle x \rangle_v = 0.47(2)$~\cite{Sutton:1991ay,Gluck:1999xe} and the lattice QCD gives $2\,\langle x \rangle_v = 0.48(4)$~\cite{Detmold:2003tm}. To match this result, we determine $\mu_0 = 0.52$\,GeV, so that $\langle x \rangle_v = 0.5$ at $\mu_0$ reduces to $\langle x \rangle_v = 0.24$ at 2 GeV by NLO DGLAP evolution. 

\begin{figure}[tbp]
\centering\includegraphics[width=\columnwidth]{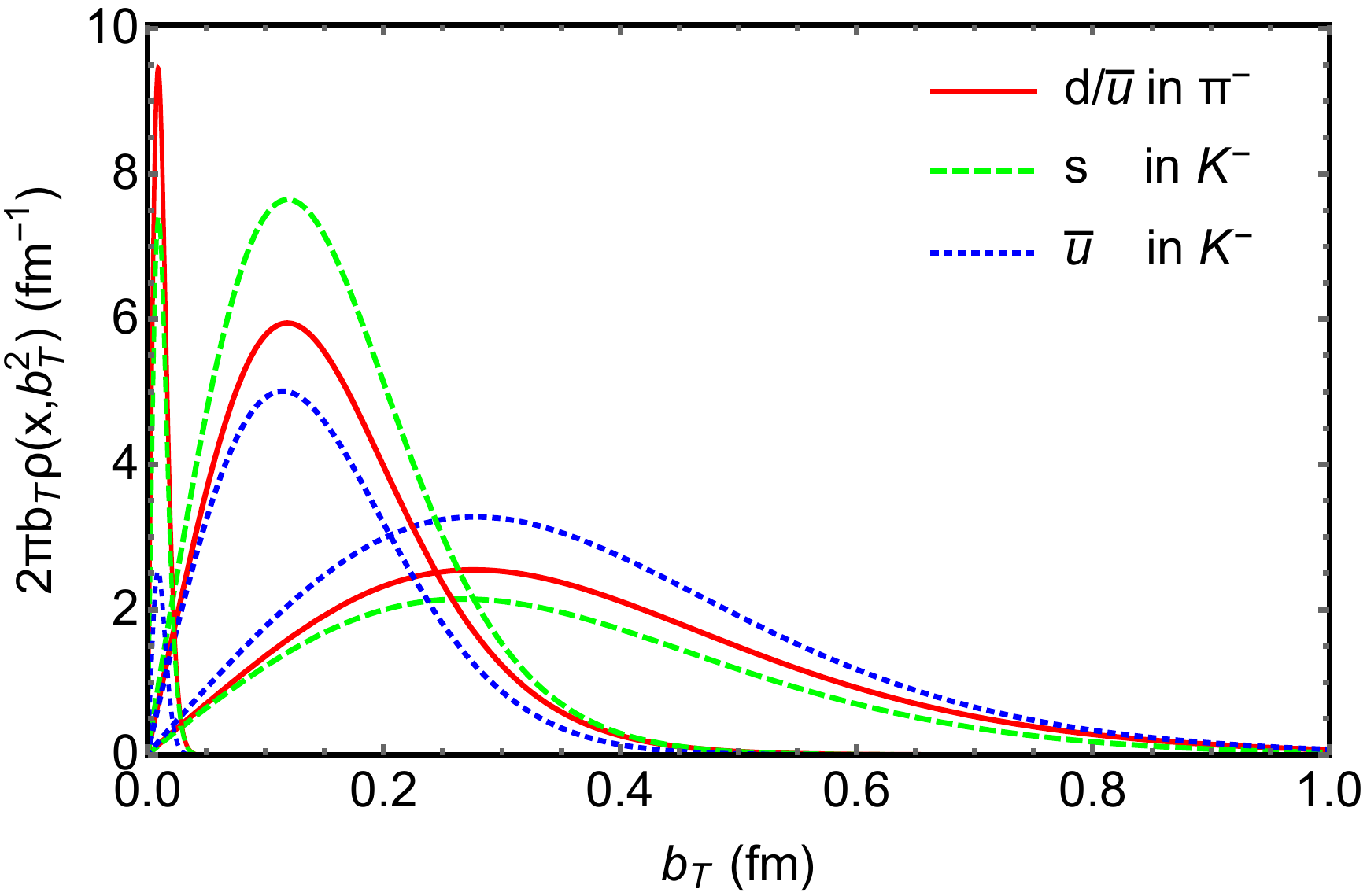} \\
\centering\includegraphics[width=0.98\columnwidth]{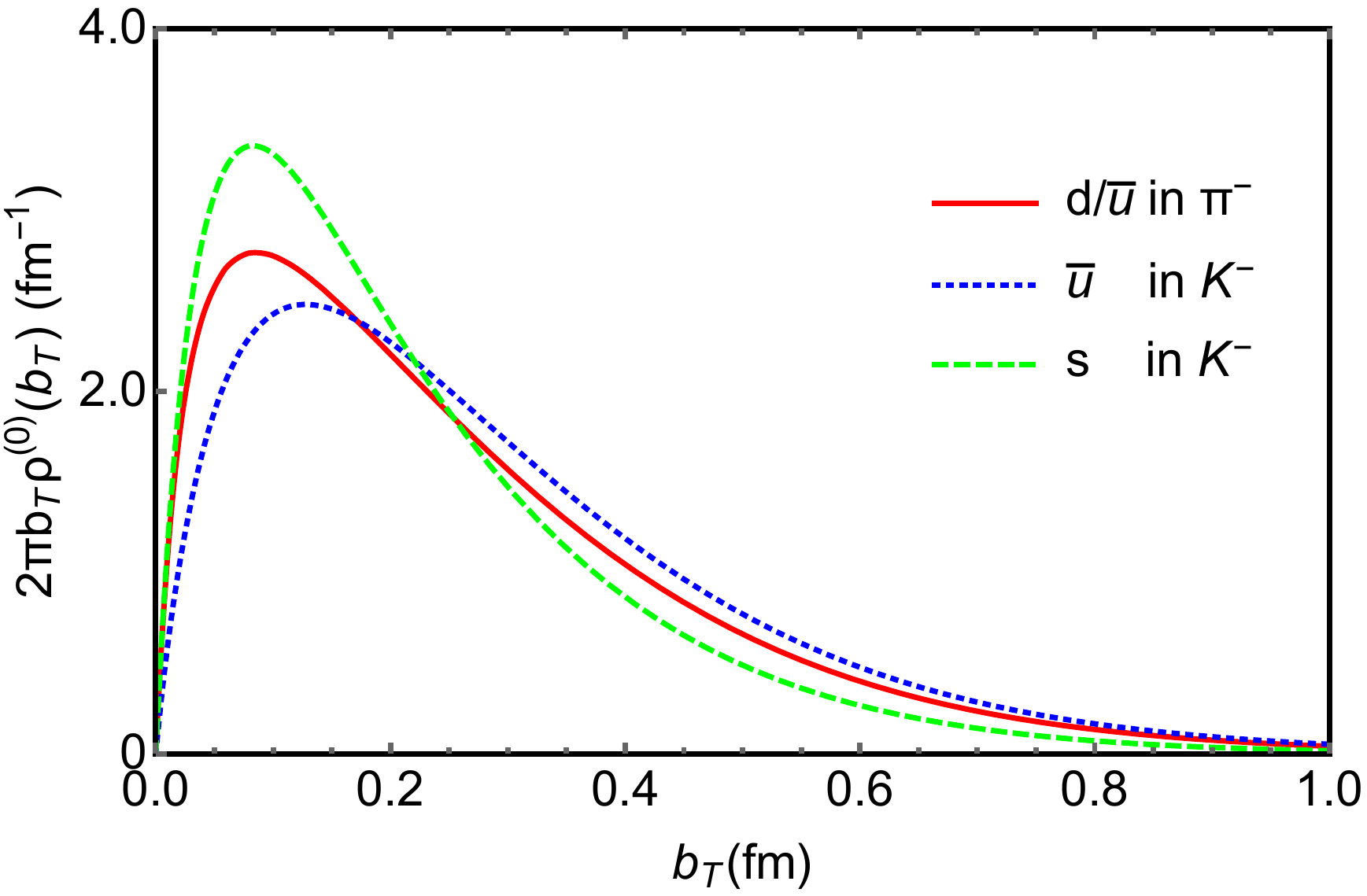} \\
\caption{\looseness=-1
{\it Upper panel:}
IPDs $\rho_M^q(x,\vect{b}_T^2)$ for the valence quarks in pion and kaon at model scale ($\mu_0 = 520$ MeV). The line styles are indicated in the plot. For each quark distribution with same line style, the three peaks from left to right correspond to $x=0.98$, $x=0.7$ and $x=0.3$ respectively.
{\it Lower panel:} The $\rho^{(0)}(\vect{b}_T^2)$ of pion and kaon as defined in Eq.~(\ref{eq:rho0}).
}
\label{fig:rhos0}
\end{figure}

The two-dimensional Fourier transform of $H_M^q(x,0,\Delta_T^2)$ gives the IPDs:
\begin{align}
\rho_M^q(x,\vect{b}_T^2) = 
\int \frac{d^2 \vect{\Delta}_T}{(2\pi)^2}\,H_M^q(x,0,-\vect{\Delta^2_T})\,e^{i \vect{b}_T \cdot \vect{\Delta}_T }, \label{eq:ipdfourier}
\end{align}
The IPDs have the interpretation of parton distributions in the transverse plane~\cite{Burkardt:2000za,Burkardt:2002hr}, with $x$ the light-cone momentum fraction and $\vect{b}_T$ the transverse separation between the active parton and the origin of transverse center of momentum $\vect{R}_T$. In the valence picture with two constituents, $\vect{R}_T = x\,\vect{r}_{T,1} + (1-x)\,\vect{r}_{T,2}$, where $\vect{r}_{T,i}$ is the transverse position of $i$th quark. The impact parameter is then $\vect{b}_{T,1}=\vect{r}_{T,1}-\vect{R}_T$. In Fig.~\ref{fig:rhos0} we plot $\rho_M^q(x,\vect{b}_T^2)$ for pion and kaon. An important observation is that as $x$ becomes larger, the width of the curves shrinks and the quark distributions are more spatially localized. When $x \rightarrow 1$, the width is vanishingly small and the quark stays near the center of transverse momentum. This can be understood since when one quark carries almost all of the light-cone  momentum (as $x\rightarrow 1$), then $\vect{R}_T \rightarrow \vect{r}_{T,1}$ and $\vect{b}_{T,1}\rightarrow 0$, namely, this quark defines the transverse center of momentum. Alternatively, if we consider the overlap representation of $\rho(x,\vect{b}_T^2)$ in terms of LFWFs in the coordinate space, that is~\cite{Kim:2008ghb,Li:2017mlw}
\begin{align}
\rho(x,\vect{b}_T^2)=\frac{1}{(1-x)^2}\sum_{\lambda_1,\lambda_2}\biggl|\tilde{\Phi}_{\lambda_1,\lambda_2}\,
\left(x,\frac{\vect{\vect{b}_T}}{1-x}\right)\biggl|^2.
\end{align}
where $\tilde{\Phi}_{\lambda_1 \lambda_2}(x,\vect{r})=\int d^2\vect{k}\,\textrm{e}^{i \vect{k}\vect{r}}\,\Phi_{\lambda_1 \lambda_2}(x,\vect{k})$,\footnote{Recall that $\Phi_{\lambda_1 \lambda_2}(x,\vect{k}_T)$ has been defined in Eq.~(\ref{eq:LFWF1}) and can be easily be related to $\psi_0(x,\vect{k}_T^2)$ and $\psi_1(x,\vect{k}_T^2)$ via comparison with Eqs.~\eqref{eq:LFWF2-0}--\eqref{eq:LFWF2-2}.} then, as $x \rightarrow 1$ the impact parameter $\vect{b}_T$ must approach zero so $\vect{b}_T/(1-x)$ doesn't go large.

Flavor symmetry breaking effects are clearly evident in Fig.~\ref{fig:rhos0}. Typically, at smaller $x$ ($x=0.3$)  there is more $\bar{u}$ quark than $s$ quark in kaon over the whole $\vect{b}_T$ range. At larger $x$ ($x=0.7$) the situation is reversed. This suggests the $s$ quark is more likely distributed near the center of kaon while the $u$ quark is more spread out.  We can also look at 
\begin{align}
\rho^{(0)}(\vect{b}_T^2) = \int_0^1 dx\ \rho(x,\vect{b}_T^2),
\label{eq:rho0}
\end{align}
which characterizes the quark density at transverse separation $\vect{b}_T$. As shown in the lower panel of Fig.~\ref{fig:rhos0}, the $s$ quark in kaon favors small $\vect{b}_T$ and $\bar{u}$ quark has a broader distribution, with $d$ quark in pion lying in between. If we look at their mean-squared $\vect{b}_T$, i.e., $\langle \vect{b}_T^2\rangle=\int d^2\vect{b}_T \vect{b}_T^2 \int_0^1 dx \rho(x,\vect{b}_T^2)$, we find  $\langle \vect{b}_T^2\rangle_{u}^\pi=0.11$\,fm$^2$, $\langle  \vect{b}_T^2\rangle_{s}^K=0.08$\,fm$^2$, and $\langle \vect{b}_T^2\rangle_{u}^K=0.13$\,fm$^2$. It's worth mentioning here that in our calculation, the current quark mass we used are $m_{u/d}^{\zeta=2\,\textrm{GeV}}=4.3$\,MeV and $m_{s}^{\zeta=2\,\textrm{GeV}}=110$\,MeV. This big mass difference gets weakened by the DCSB, and the difference in the $u/d$ and $s$ quark distributions is no longer so dramatic.  

Further, on can define the valence-like distribution $\rho^{(0)}_v(\vect{b}_T^2)=\rho^{(0)}_q(\vect{b}_T^2)-\rho^{(0)}_{\bar{q}}(\vect{b}_T^2)$, where $q$ is the active quark. Because $\rho_{\bar{q}}^{(0)}(\vect{b}_T^2)$ vanishes at the model scale in our leading Fock state calculation, then  $\rho^{(0)}_v(\vect{b}_T^2)$ is equivalent to $\rho^{(0)}_q(\vect{b}_T^2)$ plotted in Fig.~\ref{fig:rhos0}. However, it's worth mentioning that $\rho^{(0)}_v(\vect{b}_T^2)$ is independent of the renormalization scale, because DGLAP evolution conserves the quark number density at every slice of $\vect{b}_T$. Equivalently, $H(x,0,t)$ evolves independently of $t$~\cite{Burkardt:2000za}. Thus the lower panel of Fig.~\ref{fig:rhos0} can also be viewed as the valence (anti-)quark spatial distribution at any scale.

\section{ELECTROMAGNETIC AND GRAVITATIONAL FORM FACTORS\label{sec:FF}}
\begin{figure}[tbp]
\centering\includegraphics[width=\columnwidth]{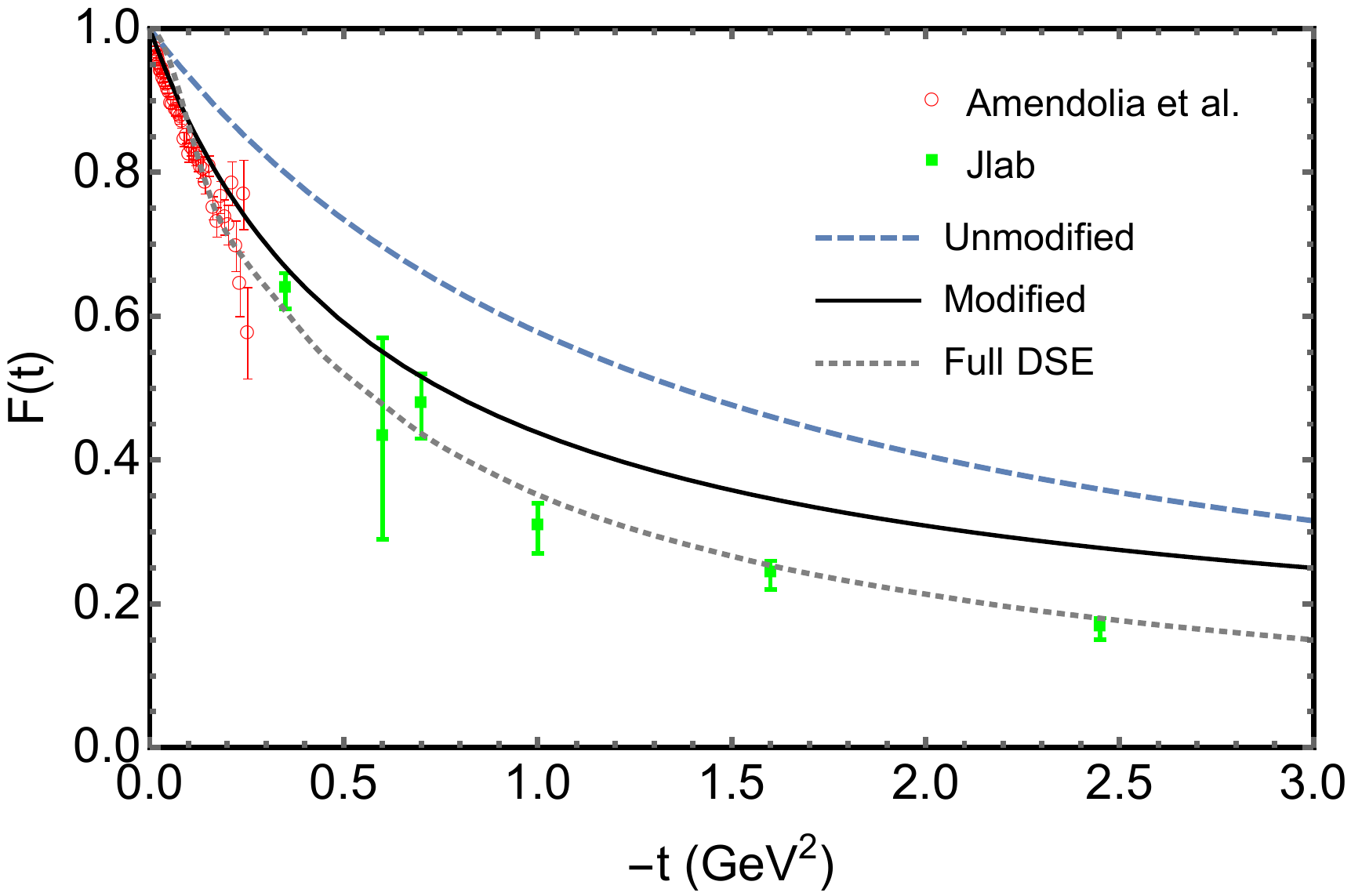} 
\caption{
Electromagnetic form factor $F(t)$ of pion in the space-like region. The data is the from NA7 Collaboration~\cite{Amendolia:1986wj} (red empty circle) and Jefferson Lab~\cite{Huber:2008id} (green filled square). The dashed (blue) curve is based on the unmodified GPD in Eq.~(\ref{eq:Hoverlap}), while the solid (black) curve uses a GPD with a dressed operator to simulate higher Fock states, see Eq.~(\ref{eq:Hmod}). The dotted curve is the full rainbow-ladder DSE result from Ref.~\cite{Maris:2000sk} that including an infinite tower of Fock states.} 
\label{fig:Ft}
\end{figure}

The electromagnetic form factors of a hadron provide important information about its spatial structure. The pion and kaon have one electromagnetic form factor defined by
\begin{align}
\sum_{q=u,d} \lf<M(p')\lf|e_q\,J_q^\mu(0)\rg|M(p)\rg> = (p'^\mu+p^\mu) F_M(t),
\end{align}
with $J_q^\mu(x)=\bar{\psi}_q(x)\gamma^\mu \psi_q(x)$ and $t = -Q^2 = (p'-p)^2$. The pion and kaon electromagnetic form factors are also given by the lowest $x$-weighted moment of their GPDs
\begin{align}
F_M(t) = \int_{-1}^1dx \lf[e_u\,H^u_M(x,\xi,t) + e_d\,H^d_M(x,\xi,t)\rg], 
\label{eq:fpi}
\end{align}
which is independent of skewness $\xi$ because of the polynomiality property of the GPDs. The result for the pion's electromagnetic form factor obtained using Eq.~\eqref{eq:fpi} is given by the dashed curve in Fig.~\ref{fig:Ft}. In general we find that our result overshoots the data for all $Q^2$, and also the full DSE calculation that uses the Bethe-Salpeter wave function and a dressed quark-photon vertex to directly calculate the pion's form factor~\cite{Maris:2000sk}. As we will explain, the origin of these discrepancies is naturally explained by the Fock state truncation and the LFWF normalization condition [see Eq.~\eqref{eq:fdef}].
 
At (very) large $Q^2$ perturbative QCD predicts that the pion's electromagnetic form factor behaves as~\cite{Lepage:1980fj}
\begin{align}
\label{eq:pionUV}
\exists\, Q_0 > & \Lambda_{\rm QCD} \; |\;   Q^2 F_\pi(Q^2) \stackrel{Q^2 > Q_0^2}{\approx} 16 \pi\, C_F\,\alpha_s(Q^2)\, w_{\pi}^2,
\end{align} 
where $w_\pi$ is the $x^{-1}$ moment of parton distribution amplitude (PDA) $w_\pi = \int_0^1dx\, x^{-1}\,\phi_\pi(x,Q^2)$, where in this case the PDA $\phi_\pi(x,Q^2)$ is normalized at the scale of $Q^2$ such that
\begin{align}
\label{eq:PDAdef}
\phi_\pi(x,Q^2)&=\int_{\vect{k}_T^2 \le Q^2}\frac{d^2 \vect{k}_T}{16 \pi^3}\  \psi_0(x,\vect{k}_T^2), \\
\int_0^1 dx\ \phi_\pi(x,Q^2)&=\frac{f_\pi}{2\sqrt{3}},  
\label{eq:PDAnorm}
\end{align}
where $f_\pi = 92.4$\,MeV is the pion's electroweak decay constant. The DSE calculation based on Eqs.~\eqref{eq:pionUV}--\eqref{eq:PDAnorm} has been presented in Ref.~\cite{Chang:2013pq} and the result is reasonable. However, the LFWF normalized by Eq.~(\ref{eq:PDAnorm}) is significantly smaller than required due to our normalization condition in Eq.~(\ref{eq:fdef}). The (very) large $Q^2$ behavior of the pion's electromagnetic form factor is dominated by the leading Fock state, and thus the deviation at large $Q^2$ can be explained by the normalization condition. Similarly, in a full calculation some of the charge of the pion with be carried by the higher Fock states, which would reduce the normalization of the leading Fock state and thereby bring our result into much better agreement with data at large $Q^2$. However, we see from Fig.~\ref{fig:Ft} that the $Q^2$ dependence of the LFWF result of the full DSE result does begin to track each other\emdash only differing by a constant normalization\emdash as $Q^2$ become large. This indicates the onset of the dominance of the leading Fock state.

The deviation in the low $Q^2$ region is also easy to understand. The normalized condition for the LFWFs is such that $F_\pi(0)=1$. However, as mentioned higher Fock states will carry some charge, which, if included, would cause a modification to the form factor at low to intermediate $Q^2$. In addition, there are important contributions that can dramatically change the charge radius but do not impact the charge. Traditionally, these are associated with vector meson dominance (VMD) contributions. VMD is associated with meson poles in the time-like region, where for the pion electromagnetic form factor the rho pole is the most important. In the LFWF approach these VMD contributions can only be obtained by including an infinite tower of Fock states. This is natural in the complete DSE calculation with a dressed quark-photon vertex, but very challenging in a rigorous light-front approach. It is therefore not possible for a leading Fock state calculation---that is intimately connected to underlying QCD dynamics---to give a good description of the electromagnetic form factor for all $Q^2$.
 
With the pion's (Breit-frame) charge radius defined by 
\begin{align}
r^2_c = - 6 \lf.\frac{\partial\, F_\pi(Q^2)}{\partial Q^2}\rg|_{Q^2 = 0},
\end{align}   
we obtain from the leading Fock state calculation $r_c=0.41$\,fm, which is significantly smaller than the experiment value of $r_c=0.67$\,fm~\cite{Beringer:1900zz}. A similar result was also found using a relativistic constituent quark model based on an effective $q\bar{q}$ Hamiltonian~\cite{Godfrey:1985xj}, where a pion charge radius of $r_c = 0.45$\,fm was found~\cite{Cardarelli:1994ix}. In this work the authors argue that the discrepancy with experiment can be corrected by taking into account the constituent quark charge radius, which is analogous to dressing the vertex as in a full DSE calculation.

In a complete DSE calculation the operator that defines the GPDs would be dressed. Such a calculation from the DSE is very difficult and beyond the scope of this work. However, we can use an analogous calculation for this dressed operator from the NJL model to obtain a qualitative measure of the impact of a dressed vertex, or equivalently higher Fock states. Using a dressed operator that defines the GPD from the NJL model~\cite{Freese:2019eww}, we find in the impulse approximation that our leading Fock state DSE result is modified such that
\begin{align}
\hs*{-2mm}H'_d(x,0,t) = H_d(x,0,t) + \delta(x)\,\tilde{F}_\rho(t)\!\int_{-1}^1\! dy\, H_{I=1}(y,0,t),
\label{eq:Hmod}
\end{align}
where
\begin{align}
H_{I=1}(x,0,t) &= H_u(x,0,t) - H_d(x,0,t),
\end{align}
and the modified GPD at zero skewness is denoted by $H'_d(x,0,t)$. Note, in the pion $H_u(x,0,t)$ can be obtained from $H_d(x,0,t)$ by charge symmetry. The second term on the right hand side of Eq.~(\ref{eq:Hmod}) comes from the dressing of the quark vertex in the impulse approximation and provides an additional contribution (see App.~\ref{sec:App-B} for details). Using $H'(x,0,t)$, we get the solid curve in Fig.~\ref{fig:Ft} and a charge radius $r_c=0.59$ fm, with the low to intermediate $-t$ region also significantly improved.

The modification term in Eq.(\ref{eq:Hmod}) has many interesting properties. For instance, its dressing function $\tilde{F}_\rho(t)$ vanishes at $t=0$, so the PDF is unchanged, i.e., $H'(x,0,0)=H(x,0,0)$. While at non-vanishing $t$, the modification term  proportional to $\delta(x)$ is infinitely negative. Its integration over $x$ yields a finite suppression to the electromagnetic form factor. In terms of the overlap representation, this correction can only be obtained by including an infinite tower of Fock states containing $\bar{q}q$ pairs. The modification in Eq.~(\ref{eq:Hmod}) brings no change to $\rho(x,\vect{b}_T^2)$ for $x>0$ and all the results in last section still hold.

The higher moments of the GPD at $\xi=0$, i.e.,
\begin{align}
\int_{-1}^1dx\, x^m\, H'^q(x,0,t) &= A^q_{m+1,0}(t)\Big|_{m\ge 1},
\label{eq:theta2}
\end{align}
are not affected by this modification term, as a consequence of the $\delta(x)$. Among these moments, $A_{2,0}^q(t)$ contributes partially to the pion's gravitational form factor $\Theta_2(t)$, defined through the matrix element of energy-momentum tensor for one-pion states~\cite{Donoghue:1991qv}
\begin{align}
\langle \pi^+(p')|\Theta^{\mu\nu}(0)|\pi^+(p)\rangle&=\frac{1}{2}[P^\mu P^\nu\Theta_2(t) \nonumber \\
&\hs*{8mm}
+(g^{\mu \nu}q^2-q^\mu q^\nu)\Theta_1(t)].
\end{align}
with $P=p+p'$, $q=p'-p$ and $t=q^2$. The form factor $\Theta_2(t)$ is scale independent, while its individual quark contributions $A_{2,0}^q(t)$ evolve with scale. At the low model scale, the valence picture gives $\Theta_2(t) = \sum_q A_{2,0}^q(t)$. As the scale increases, $A_{2,0}^q(t;\mu)$ evolves accordingly to the evolution of the GPD. 

\begin{figure}[tbp]
\centering\includegraphics[width=\columnwidth]{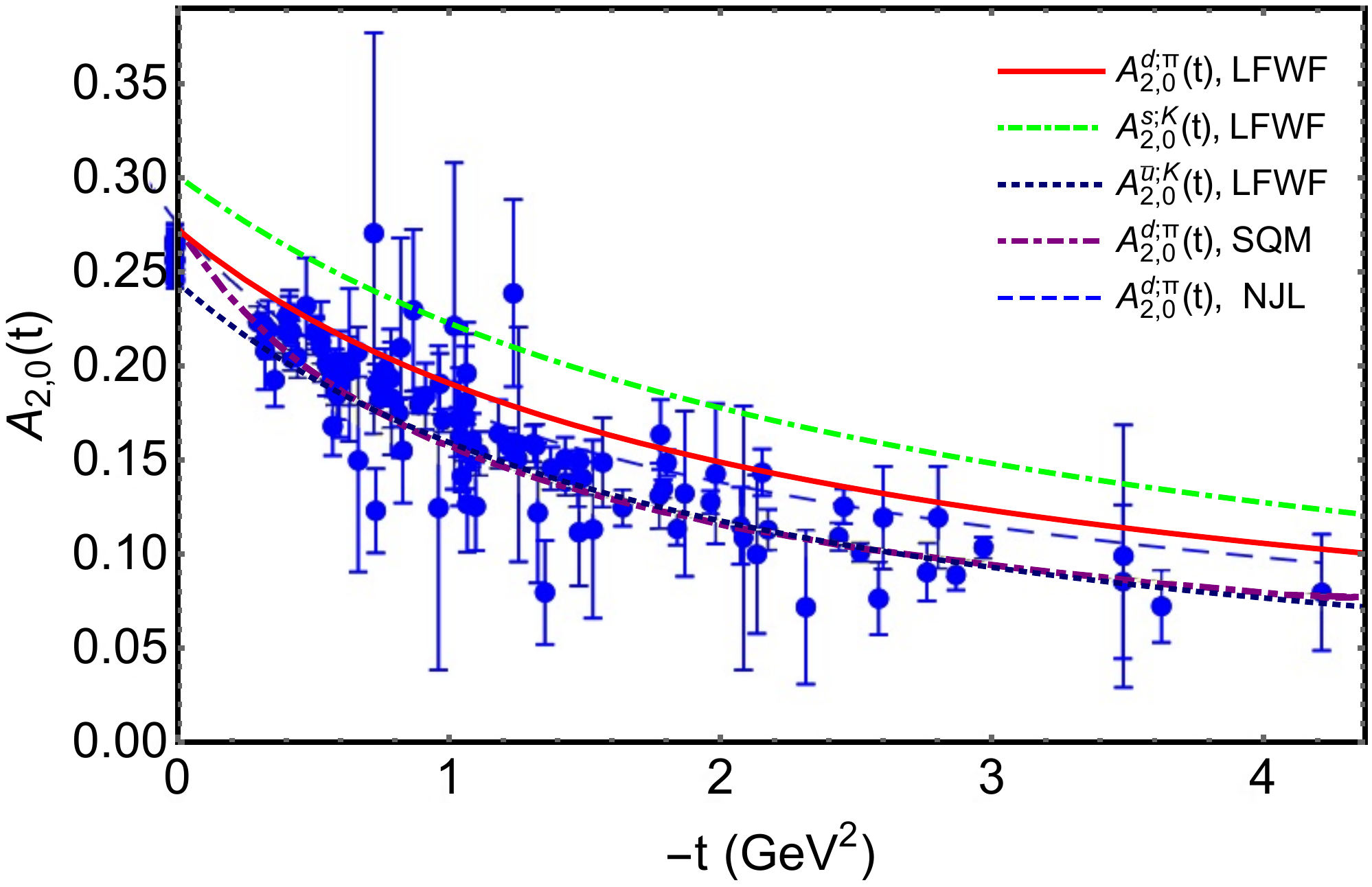} 
\caption{\looseness=-1 The quark part of gravitational form factor $A_{2,0}^q(t)$ in pion and kaon. The solid, dash-dash-dotted and dotted curves are obtained by our DSEs-based LFWFs. All the other curves and data are taken from~\cite{Broniowski:2008hx} .The dot-dashed curve is the spectral quark model prediction and the dashed curve is by NJL model with the Pauli-Villars (PV) regularization. The data is from lattice QCD~\cite{Brommel:2007zz}. 
\label{fig:GFF}}
\end{figure}

In Fig.~\ref{fig:GFF} we show pion's $A_{2,0}^{d;\pi}(t)$ (solid red curve) at the scale of 2\,GeV and the curve lies within the lattice simulation data. It is closer to the NJL model result (blue dashed)~\cite{Broniowski:2008hx} than to the spectral quark model~\cite{Broniowski:2008hx}. We have illustrated the kaon GFFs $A_{2,0}^{\bar{u};K}(t)$ and $A_{2,0}^{s;K}(t)$ as well. 

A light-cone energy radius can be defined in relation to the gravitational form factor $A_{2,0}(t)$, and is given by~\cite{Freese:2019bhb}
\begin{align}
\lf< r_{E,{\rm LC}}^2 \rg> = -4\,\lf.\frac{\partial\, A_{2,0}(Q^2)}{\partial Q^2} \rg|_{Q^2=0},
\end{align}
which can be contrasted with an analogous light-cone charge radius defined by $\big< r_{c,{\rm LC}}^2 \big> = -4\lf.\partial\, F(Q^2)/\partial Q^2 \rg|_{Q^2=0}$. For the pion we find $r_{c,{\rm LC}}^{u,\pi} = 0.331\,$fm and $r_{E,{\rm LC}}^{u,\pi} = 0.185$, meaning the energy radius is about 56\% smaller that the light-cone charge radius. Both these radii will be impacted by higher Fock states, however, based on vector meson dominance the light-cone charge radius will increase more because it is impacted by the $\rho$ meson pole whereas the light-cone energy radius is impacted by spin-2 mesons which are much heavier and further from $Q^2 = 0$. Therefore, we predict that $r_{E,{\rm LC}}^{u,\pi}/r_{c,{\rm LC}}^{u,\pi} = 0.56$ is an upper bound on this ratio. For the kaon we find light-cone charge radii of $r_{c,{\rm LC}}^{u,K} = 0.358\,$fm and $r_{c,{\rm LC}}^{s,K} = 0.281\,$fm, and light-cone energy radii of $r_{E,{\rm LC}}^{u,K} = 0.192\,$fm and $r_{E,{\rm LC}}^{s,K} = 0.173\,$fm. In each case the $s$ quark has a smaller extent than the $u$ quark.

\begin{figure}[tbp]
\centering\includegraphics[width=\columnwidth]{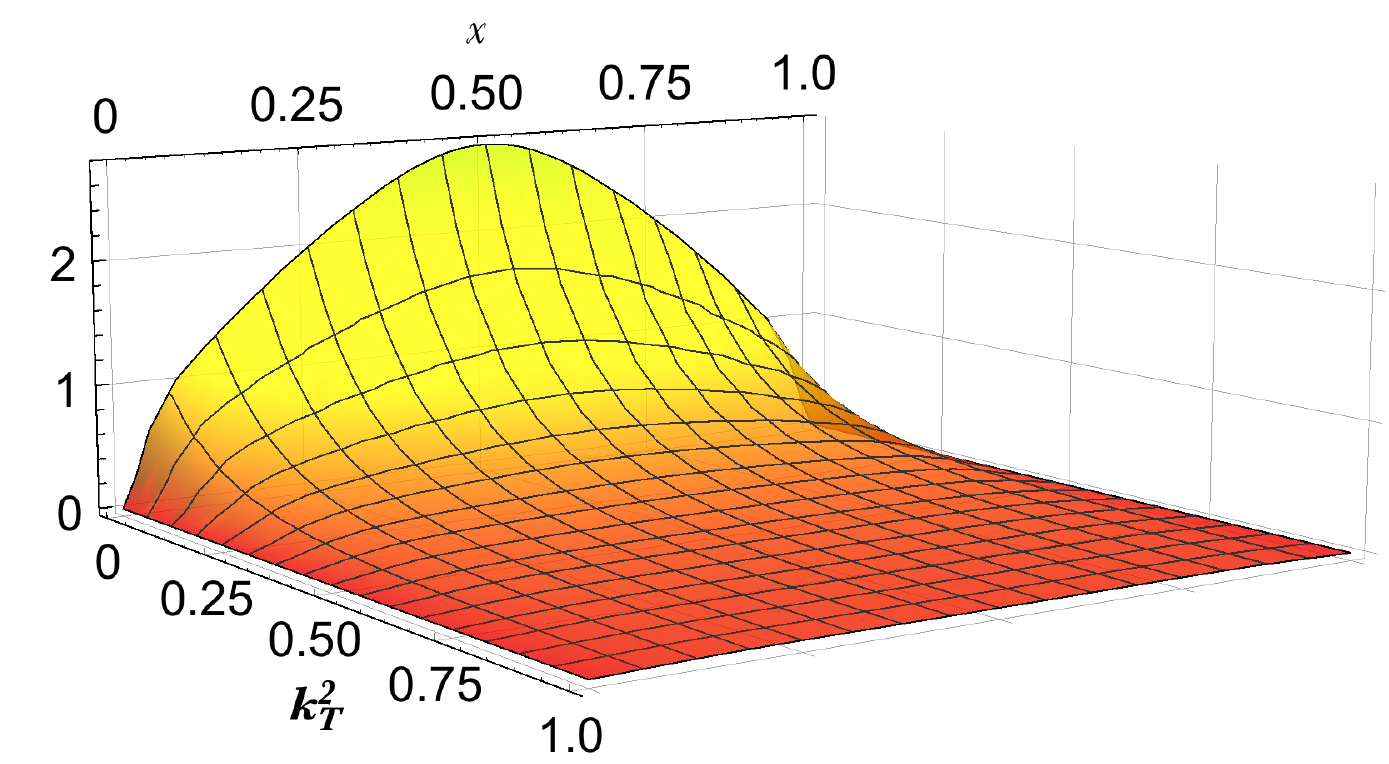} \\
\centering\includegraphics[width=0.98\columnwidth]{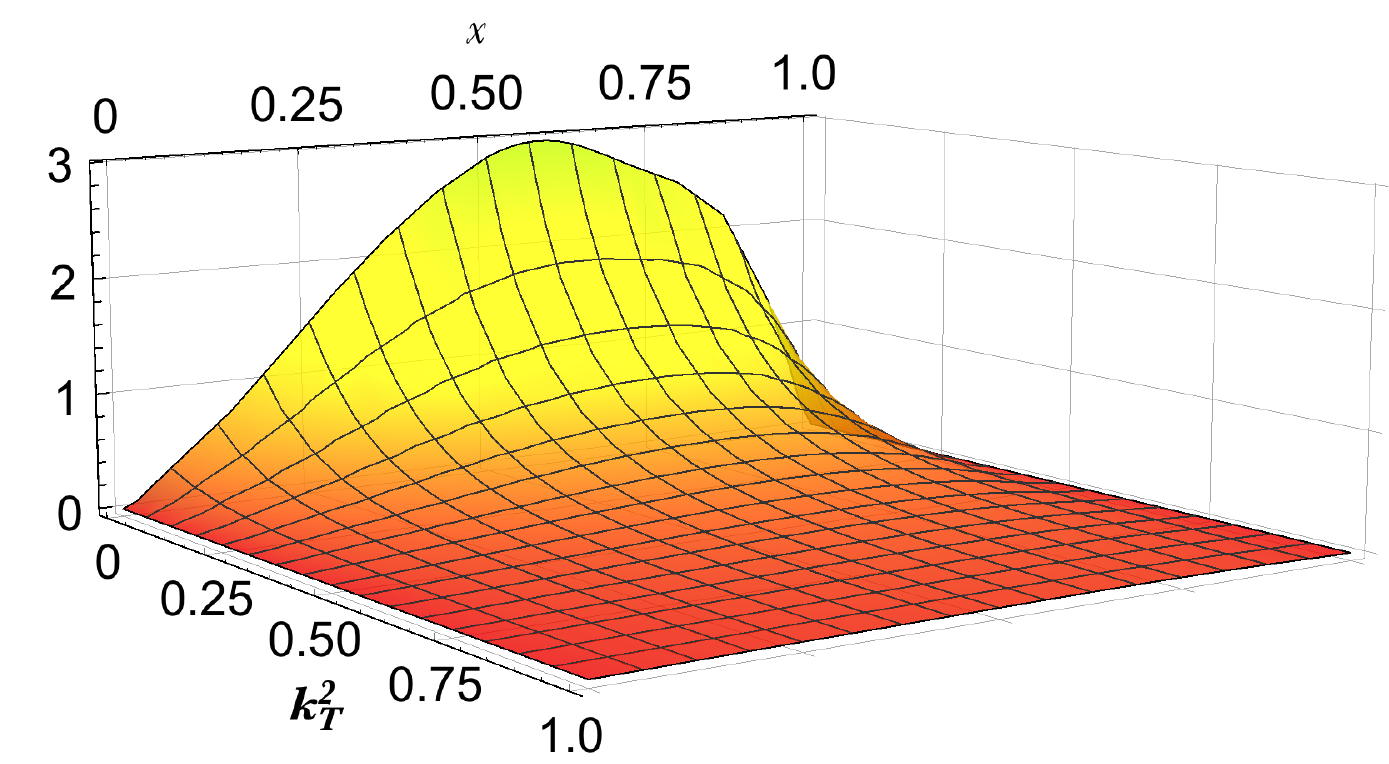} \\
\caption{The unpolarized TMD $f^d_{1;\pi}(x,\vect{k}_T^2)$ of pion (upper panel)
and $f^s_{1;K}(x,\vect{k}_T^2)$ of kaon (lower panel).}
\label{fig:tmds}
\end{figure}

\section{TRANSVERSE MOMENTUM DEPENDENT PARTON DISTRIBUTION FUNCTIONS\label{sec:TMD}}
The unpolarized leading-twist TMD is defined as
\begin{align} 
f_{1}(x,\vect{k}_T^2)&=\int\frac{d \xi^-d^2\vect{\xi}_T}{(2 \pi)^3}\ e^{i(\xi^-k^+-\vect{\xi}_T\cdot \vect{k}_T)}\nonumber\\
&\hspace{28mm}
\langle P|\bar{\psi}(0)\gamma^+\psi(\xi^-,\vect{\xi}_T)|P\rangle,
\end{align}
with the gauge link omitted. In terms of the leading Fock state LFWFs the TMD reads~\cite{Pasquini:2014ppa}
\begin{align}
f^q_1(x,\vect{k}_T^2) = \frac{1}{(2 \pi)^3} 
\left[\lf|\psi_0(x,\vect{k}_T^2)\rg|^2 + \vect{k}_T^2 \lf|\psi_1(x,\vect{k}_T^2)\rg|^2\right], 
\label{eq:tmd}
\end{align}
which should be associated with an initial result at a low scale of $\mu_0=520$\,MeV, just as in the GPD case. We plot the unpolarized TMD for the pion and kaon ($s$ quark) in Fig.~\ref{fig:tmds}. The $\bar{u}$ TMD in kaon can be simply obtained from $s$ quark distribution by momentum conservation, i.e., $f^{\bar{u}}_{1\;K}(x,\vect{k}_T^2) = f^{s}_{1\;K}(1-x,\vect{k}_T^2)$. The distribution closely resembles the profile of LFWFs. Which suggests that in a purely valence quark picture the quarks are most likely to carry around half of the parent hadron's light-cone momentum with a small intrinsic transverse momentum. 

\begin{figure}[tbp]
\centering\includegraphics[width=\columnwidth]{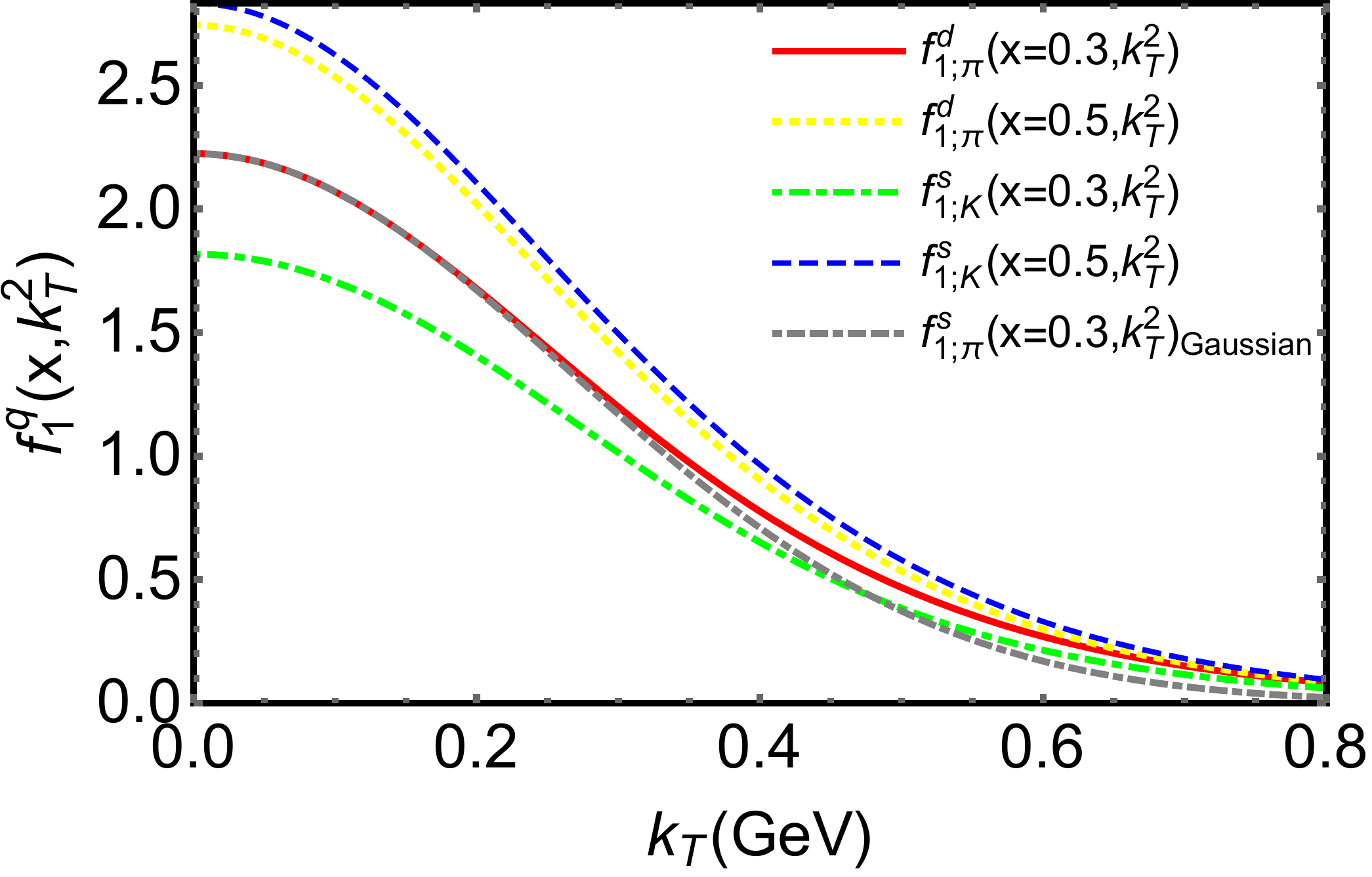} \\
\caption{
The $|\vect{k}_T|$-dependence of pion's and kaon's unpolarized TMD at certain $x$ values. The line styles are indicated in the plot and further explained in the text.}
\label{fig:gaussian}
\end{figure}

The transverse momentum dependence of the TMD has long been of great interest, and in Fig.~\ref{fig:gaussian} we illustrate our results for fixed values of $x$. Our results decrease with increasing $|\vect{k}_T|$, being concave at low $|\vect{k}_T|$ and becoming convex as $|\vect{k}_T|$ increases. The inflection point is around 300\,MeV. Phenomenologically, postulating a Gaussian $|\vect{k}_T|$-dependence is popular and Gaussian-based models successfully describe much of the existing data~\cite{DAlesio:2004eso,Anselmino:2005nn,Collins:2005ie,Schweitzer:2010tt,Aybat:2011zv,Wang:2017zym,Bacchetta:2017gcc}. The gray dot-dash-dash curve is a Gaussian function $f_{\textrm G}(\vect{k}_T^2)=N e^{-\vect{k}_T^2/\langle \vect{k}_T^2 \rangle}$ employed to fit $f^d_{1;\pi}(x=0.3,\vect{k}_T^2)$ at low $|\vect{k}_T|$. One can see the fit is good up to around 300\,MeV, i.e., it describes well the intrinsic transverse momentum dependence. For large $|\vect{k}_T|$ the Gaussian form would inevitably fail since $f_1^q(x,\vect{k}_T^2)\sim 1/\vect{k}_T^4$ with our LFWFs.

The $x$-dependence and $\vect{k}_T$-dependence in our TMDs are not factorizable, except for at very large $\vect{k}_T^2$. For instance, in Fig.~\ref{fig:gaussian}, the $\langle \vect{k}_T^2\rangle$ is respectively 0.14\,GeV$^2$ and 0.13\,GeV$^2$ when fitting $f^d_{1;\pi}(x=0.3,\vect{k}_T^2)$ and $f^d_{1;\pi}(x=0.5,\vect{k}_T^2)$ to the Gaussian form $f_{\textrm G}(\vect{k}_T^2)=N e^{-\vect{k}_T^2/\langle \vect{k}_T^2 \rangle}$. In recent years,  phenomenological studies of TMDs have appreciated the $x$-dependence of the $|\vect{k}_T|$ behavior and build it this into their parameterizations at the low initial scale~\cite{Bacchetta:2017gcc,Scimemi:2017etj}. In this sense, our result shows qualitative agreement. Note that TMD evolution also generates significant $x$-dependence in the $|\vect{k}_T|$ behavior, as has been shown in~Refs.~\cite{Shi:2018zqd,Bacchetta:2017vzh}.  Finally, we report that for the $s$ quark in kaon, $\langle \vect{k}_T^2\rangle$ is  respectively 0.155\,GeV$^2$ and 0.134\,GeV$^2$ when fitting $f^s_{1;K}(x=0.3,\vect{k}_T^2)$ and $f^s_{1;K}(x=0.5,\vect{k}_T^2)$ to the Gaussian form. This is slightly larger than the $\bar{u}$ or $d$ quarks in pion, and is a measure of SU(3) flavor symmetry breaking. 

\section{CONCLUSION\label{sec:con}}
By projecting the mesons' covariant Bethe-Salpeter wave functions onto the light front, we calculate the leading Fock state LFWFs of the pion and kaon. The kaon's LFWFs based on a DSE approach are given for the first time. These LFWFs are significantly enhanced at low $|\vect{k}_T|$ and exhibit the perturbative QCD power law behavior at large $|\vect{k}_T|$. SU(3) flavor symmetry breaking is revealed in the kaon's LFWFs. We also observe a sizable contribution from the spin-parallel LFWF, suggesting an important role played by the $p$-wave component in the pion and kaon as relativistic composite particles. 

We employ the light front overlap representation given in Eq.~(\ref{eq:Hoverlap}) to study the GPDs at zero skewness $H(x,\xi=0,t)$ for the pion and kaon, and using the IPDs $\rho(x,\vect{b}_T^2)$, we determine the spatial distribution of the valence quarks. On the light front the quarks with larger light-cone momentum fraction $x$ are generally less spread out in the spatial impact parameter $\vect{b}_T$. After integration over $x$, we find the heavier quarks are more concentrated at the center of meson, e.g., the $s$ quark spatial distribution in $K^-$ is narrower than the $\bar{u}$ quark. 

Shortcomings in a leading Fock state truncation are exposed in the pion's electromagnetic form factor. An attempt is made to overcome these issues using a dressing of the operator that defines the GPD obtained from the NJL model. The pion and kaon electromagnetic and gravitational form factor then show reasonable agreement with available experimental and lattice data.

Finally, we give the unpolarized TMDs of pion and kaon. The phenomenologically popular Gaussian-like $\vect{k}_T$-dependence is observed in our result for intrinsic $|\vect{k}_T|$, but violated at medium and large $|\vect{k}_T|$. It is also observed that the $|\vect{k}_T|$ behavior in our TMD is slightly $x$-dependent, suggesting an unfactorizable $x$- and $\vect{k}_T$-dependence. In addition, the valence quarks in kaon have a broader transverse momentum distribution, as a consequence of SU(3) flavor symmetry breaking. Starting with the DSE Bethe-Salpeter wave function we have therefore obtain comprehensive insights into the pion and kaon valence quark imaging in both position and momentum space.

\begin{acknowledgments}
Chao Shi gives special thanks to C\'edric Mezrag for valuable suggestions in completing this work.
This work is supported by the National Natural Science Foundation of China (under Grant No. 11905104) and starting grant of Nanjing University of Aeronautics and Astronautics (under Grant No. 1006-YAH20009), by the U.S. Department of Energy, Office of Science, Office of Nuclear Physics, contract no.~DE-AC02-06CH11357; the Laboratory Directed Research and Development (LDRD) funding from Argonne National Laboratory, project no.~2016-098-N0 and project no.~2020-020; and the DOE Office of Science Graduate Student Research (SCGSR) Program.
\end{acknowledgments}

\appendix
\section{Parameterization of $\vect{S(k)}$ and $\vect{\Gamma(k;P)}$}\label{sec:para}

\begin{table}[b]
\caption{Representation parameters. \emph{Upper panel}: Eq.~(\ref{eq:spara}) -- the pair $(x,y)$ represents the complex number $x+ i y$.  \emph{Lower panel}: Eqs.~\eqref{eq:gammapara}--\eqref{eq:rho}.  (Dimensioned quantities are given in GeV).
\label{Table:parameters}
}
\begin{center}
\begin{tabular*}
{\hsize}
{
@{\extracolsep{0ptplus1fil}}
c@{\extracolsep{0ptplus1fil}}
c@{\extracolsep{0ptplus1fil}}
c@{\extracolsep{0ptplus1fil}}
c@{\extracolsep{0ptplus1fil}}
c@{\extracolsep{0ptplus1fil}}}\hline
  & $z_1$ & $m_1$  & $z_2$ & $m_2$ \\
$u$ &    $(0.44, 0.28)$ & $(0.46, 0.18)$ & $(0.12, 0.00)$ & $(-1.31, -0.75)$ \\
$s$ &    $(0.43, 0.30)$ & $(0.55, 0.22)$ & $(0.12, 0.11)$ & $(-0.83, 0.42)$ \\
\hline
\end{tabular*}

\begin{tabular*}
{\hsize}
{
l@{\extracolsep{0ptplus1fil}}
l@{\extracolsep{0ptplus1fil}}
c@{\extracolsep{0ptplus1fil}}
c@{\extracolsep{0ptplus1fil}}
c@{\extracolsep{0ptplus1fil}}
c@{\extracolsep{0ptplus1fil}}
c@{\extracolsep{0ptplus1fil}}
c@{\extracolsep{0ptplus1fil}}
c@{\extracolsep{0ptplus1fil}}
c@{\extracolsep{0ptplus1fil}}}\hline
   & $U_1$ & $U_2$ & $U_3$ &$n_1$ &$n_2$ &$n_3$ & $\sigma^i_1$ & $\sigma^i_2$ & $\Lambda$ \\[0.7ex]\hline
 E$_\pi$ & $2.76$ & $-1.84$ & $0.04$ & $4$
    & $5$ & $1$&0.0 &2.2&1.41 \\
   F$_\pi$ & $1.46$ & $-0.97$ & $0.006$ &$4$
    & $5$ & $1$&0.0&-0.5& 1.13   \\
 E$_K$ & $2.98$ & $-2.0$ & $0.025$ & $4$
    & $5$ & $1$&-0.4 &1.0&1.35 \\
   F$_K$ & $0.86$ & $-0.30$ & $0.004$ &$4$
    & $6$ & $1$&-0.4&-1.0& 1.20  \\\hline
\end{tabular*}
\end{center}
\end{table}

%
To aid the calculation of the moments $\left< x^m \right>_{l_z}$ we use an accurate parametrization of numerical solutions to the gap and BSEs in the DCSB-improved truncation to the DSEs~\cite{Chang:2010hb,Chang:2013pq}. Solutions of the DSE-BSE with the so-called DCSB-improved kernel are available within the literature, both for the pion and the kaon~\cite{Chang:2013pq,Shi:2014uwa}. In this work, we employ these results and their available parameterization, that we remind the reader of and slightly modify. The quark propagator $S(k)$ is written as the sum of pairs of complex conjugate poles:
\begin{align}
\label{eq:spara}
S(k)=\sum_{i=1}^{n}\left [ \frac{z_i}{i \sh{k}+m_i}+\frac{z^*_i}{i \sh{k}+m^*_i} \right ],
\end{align}
with $n=2$. The pseudo-scalar Bethe-Salpeter amplitude $\Gamma(k;P)$ can be generally decomposed as 
\begin{align}
\label{eq:gammapara}
\Gamma(k;P) &= \gamma_5 \Big[i E(k;P)+\sh{P} F(k;P)\nonumber \\
&\hs{22mm}
+\sh{k}\,G(k;P)+[\sh{P},\sh{q}]\,H(k;P)\Big].
\end{align}
We employ the dominant terms $E(k;P)$ and $F(k;P)$, which are parameterized by:
\begin{align}
\label{eq:fpara}
{\cal F}(k;P)&=\int_{-1}^1 d\alpha \rho_{i}(\alpha)\bigg[\frac{U_1 \Lambda^{2 n_1}}{(k^2+\alpha k\cdot P+\Lambda^2)^{n_1}}\nonumber \\ 
&\hs{30mm}
+\frac{U_2 \Lambda^{2 n_2}}{(k^2+\alpha k\cdot P+\Lambda^2)^{n_2}}\bigg]\nonumber\\ 
&+\int_{-1}^1 d\alpha \rho_u(\alpha)\frac{U_3 \Lambda^{2 n_3}}{(k^2+\alpha k\cdot P+\Lambda^2)^{n_3}}, \allowdisplaybreaks[2]\\
\rho_i(\alpha)&=\frac{1}{\sqrt{\pi}}\frac{\Gamma(3/2)}{\Gamma(1)}\Big[C_0^{(1/2)}(\alpha) \no \\
&\hs{22mm}
+ \sigma^i_1 C_1^{(1/2)}(\alpha) +\sigma^i_2 C_2^{(1/2)}(\alpha)\Big],
\label{eq:rho}
\end{align}
where $\rho_u(\alpha)=\frac{3}{4}(1-\alpha^2)$ and  \{$C_n^{(1/2)}, n=0,1,...,\infty$\} are the Gegenbauer polynomials of order $1/2$. The value of the  parameters are listed in Tab.~\ref{Table:parameters}. The outgoing quark and anti-quark in the meson carry momentum $k+P/2$ and $k-P/2$ respectively, so ${\cal F}(k;P)$ is even in $k \cdot P$ due to charge parity. 

\begin{figure}[bp]
\centering\includegraphics[width=0.7\columnwidth]{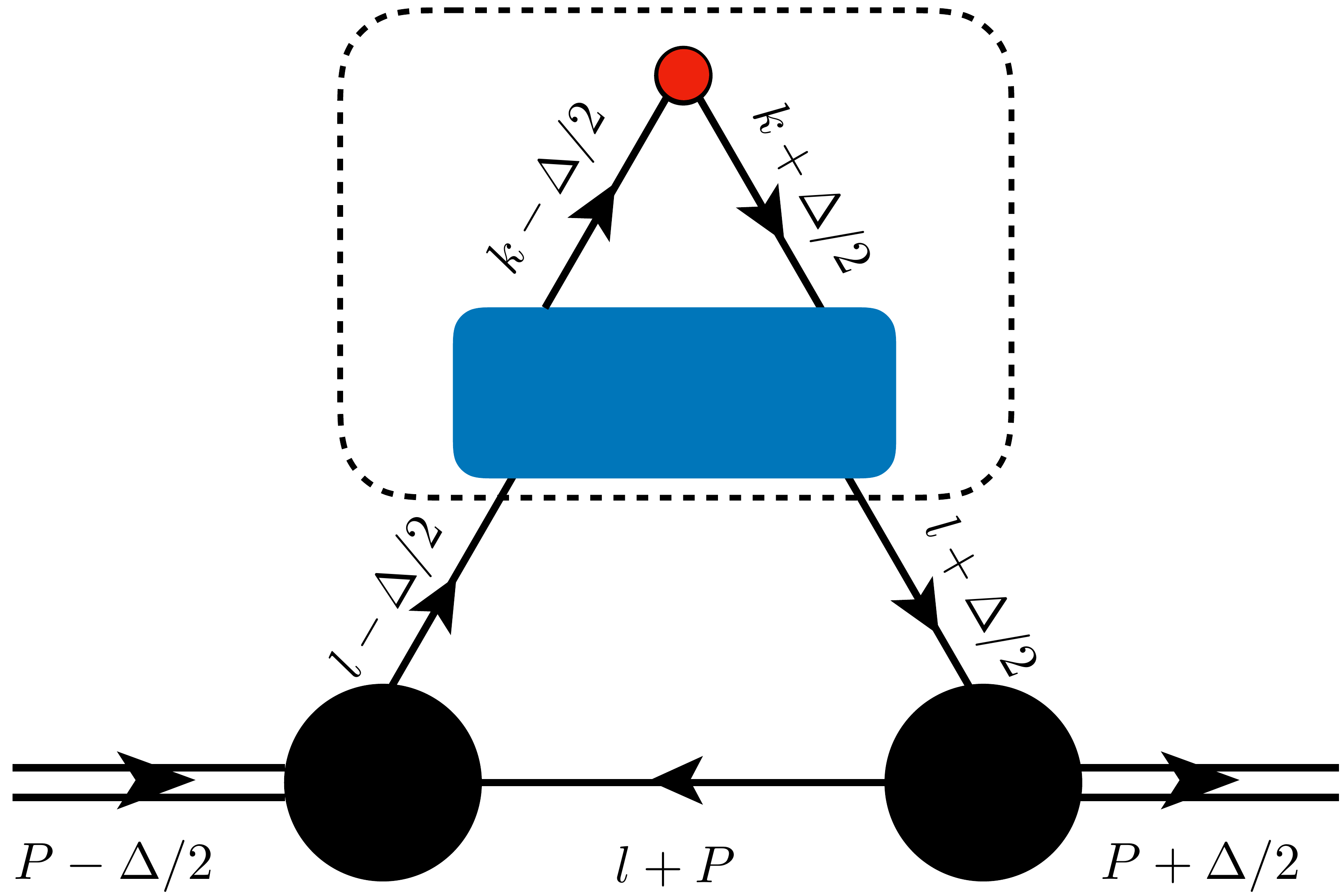} 
\caption{Triangle diagram (impulse approximation) for $H(x,\xi,t)$. The dash line boxed area represents the fully dressed vertex $\Gamma \cdot n$. Lines with arrows indicate dressed quark propagators $S$ and the black blob represents the pion's Bethe-Salpeter amplitude $\Gamma_\pi$. 
\label{fig:TriAng}}
\end{figure}

\section{Dressed GPD operator at zero skewness in NJL model\label{sec:App-B}}
The modified GPD given in Eq.~(\ref{eq:Hmod}) is important in validating our valence picture of pion concerning the pion's charge radius. Here we give a quick sketch on how it is obtained in the NJL model, hence list only the basic idea and important steps/results. Within the impulse approximation, the pion's GPD in the NJL model can be calculated as 
\begin{align}
H'_{I=0,1}(x,\xi,t)=\int \frac{d^4l}{(2\pi)^4}\ \textrm{Tr}[\bar{\Gamma}_\pi~S ~\Gamma \cdot n ~S~ \Gamma_\pi~S], 
\label{eq:TriAng}
\end{align} 
with momentum assigned in Fig. ~\ref{fig:TriAng}. Here we consider the GPD of isospin 0 or 1, defined as  $H'_{I=0}=H'_u+H'_d$ and $H'_{I=1}=H'_u-H'_d$. Flavor matrices are implicitly embedded in the elements $S$, $\Gamma_\pi$ and $\Gamma \cdot n$. The notation $\Gamma \cdot n$ represents the dash line boxed area. We denote the $\Gamma \cdot n$ as a violet blob in other diagrams. It satisfies the inhomogeneous  Bethe-Salpeter equation:
\begin{align}
\vcenter{\hbox{\includegraphics[width=0.4\textwidth]{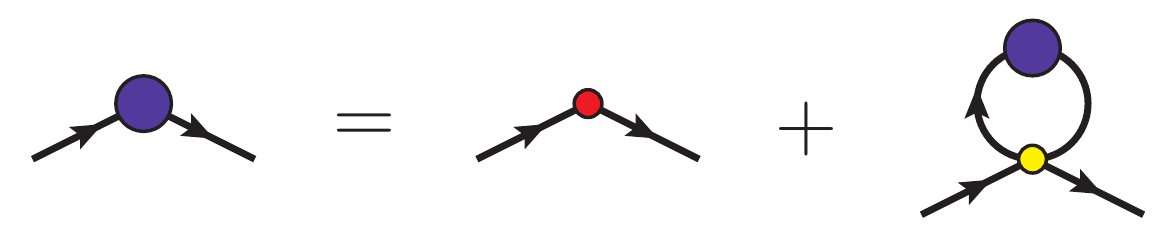}}}
\end{align}
\noindent Here the red blob is the bare vertex
\begin{align}
 \hspace{-5mm} \vcenter{ \hbox{ \includegraphics[scale=0.5]{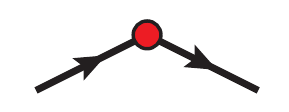} } }\hspace{-2mm}=
 \int \frac{d^4k}{(2\pi)^4} \delta^4(k-l) \left\{\slashed{n} \delta\big(n\cdot[xP-k]\big)\right\} \otimes
\begin{bmatrix} \tau_0 \\ \tau_3 \end{bmatrix}.
\end{align}
Note the first Dirac delta ensures $k\pm \Delta/2=l\pm \Delta/2$ at leading truncation.  The matrices $\tau_0$ or $\tau_3$ are for isospin 0 or 1, respectively. To solve for $\Gamma \cdot n$, one can formally sum the series 
\begin{align}
\vcenter{\hbox{\includegraphics[width=0.4\textwidth]{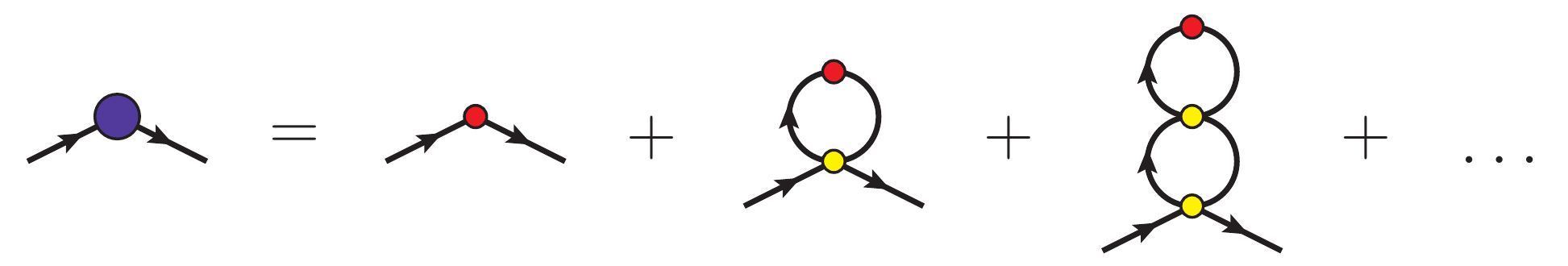}}}
\end{align}
Its sub-leading term
\begin{align}
  B_{I=0,1}(x,\xi,t)
  =
  \vcenter{ \hbox{ \includegraphics[scale=0.5]{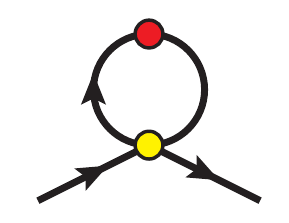} } }
\end{align}
can be evaluated via Mellin moments, i.e.,
\begin{align}
  \int \mathrm{d}x\, x^s
  B_I(x,\xi,t)
  &= \sum_\Omega 2iG_\Omega
  \hspace{-1.5mm}\int  \hspace{-1.5mm}\mathrm{d}x\, x^s
  \hspace{-2mm}\int  \hspace{-2mm} \frac{\mathrm{d}^4k}{(2\pi)^4}  \delta\big(n\cdot[xP-k]\big) \nonumber \\
&\hspace{-3mm}  \mathrm{Tr} \left[
    S\left(k+\frac{\Delta}{2}\right) \slashed{n}
    \left\{ \begin{matrix} \tau_0 \\ \tau_3 \end{matrix} \right \}
    S\left(k-\frac{\Delta}{2}\right) \Omega
    \right]
  \Omega,
\end{align}
where $\Omega$ denotes any of the five Dirac/isospin structures appearing
in the NJL model Lagrangian.\footnote{The $q\bar{q}$ interaction kernel in NJL model is given by:
\begin{align}
  K_{\alpha\beta,\gamma\delta}
  &=
  2i G_\pi \left[
    (1)_{\alpha\beta}(1)_{\gamma\delta}
    - (\gamma_5\tau_i)_{\alpha\beta} (\gamma_5\tau_i)_{\gamma\delta}
    \right]
 \nonumber \\
& - 2i G_\rho \left[
    (\gamma_\mu\tau_i)_{\alpha\beta}(\gamma^\mu\tau_i)_{\gamma\delta}
    + (\gamma_\mu\gamma_5\tau_i)_{\alpha\beta}(\gamma^\mu\gamma_5\tau_i)_{\gamma\delta}
    \right]
 \nonumber \\
& -2i G_\omega (\gamma_\mu)_{\alpha\beta}(\gamma^\mu)_{\gamma\delta}.
\end{align}}
At $\xi=0$ one finds
\begin{align}
  (P\cdot n)
  \int \mathrm{d}x\, x^s
  B_{I=0,1}(x,0,t) &=
\left\{
    \begin{array}{ll}
      - 2G_{\omega,\rho} \Pi_{VV}(t) \slashed{n} & : s = 0
      \\
      0 & : s \geq 1
    \end{array} \right.
\end{align}
which uniquely determines the result for $B_I$ to be:
\begin{align}
  B_{I=0,1}(x,0,t) &= 
  - 2G_{\omega,\rho} \Pi_{VV}(t) \frac{\delta(x)}{(P\cdot n)} \slashed{n}
  .
\end{align}
One can calculate the rest terms analogously and their summation gives the overall dressed quark correlator:
\begin{align}
  \vcenter{ \hbox{ \includegraphics[scale=0.5]{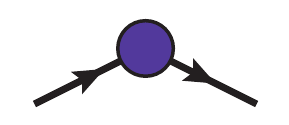} } }\biggl|^{I=0,1}_{\xi=0}
&  =
    \left\{ \int \frac{d^4k}{(2\pi)^4} \delta^4(k-l) \delta(n\cdot[xP-k])\right.
\nonumber   \\
&\hs{-5mm}
\left.- \frac{
      2G_{\omega,\rho} \Pi_{VV}(t)
    }{
      1 + 2G_{\omega,\rho} \Pi_{VV}(t)
    }
    \frac{\delta(x)}{(P\cdot n)}
\right\}    \slashed{n} \otimes \begin{bmatrix} \tau_0 \\ \tau_3 \end{bmatrix}.
\label{eq:Gamma}
\end{align}
Putting Eq.~(\ref{eq:Gamma}) back into Eq.~(\ref{eq:TriAng}), the first term in the braces gives  the bare vertex contribution
\begin{align}
H_{I=0,1}(x,0,t)&=\int \frac{d^4k}{(2\pi)^4}\delta(xP\cdot n-l\cdot n)\nonumber \\
&\hs*{15mm}
\textrm{Tr}\left\{\bar{\Gamma}_\pi~S ~\slashed{n}\otimes 
\begin{bmatrix} \tau_0 \\ \tau_3 \end{bmatrix}
~S~ \Gamma_\pi~ S\right\}. \label{eq:TriAng0}
\end{align} 
The second term has $x$ dependence factored out and one easily finds its contribution is proportional to the lowest moment of Eq.~(\ref{eq:TriAng0}). Finally we have
\begin{align}
H'_{I}(x,0,t)&=H_{I}(x,0,t)\nonumber\\
&+\frac{-2 G_{\omega,\rho}\Pi_{VV}(t) \delta(x)}{1+2 G_{\omega,\rho}\Pi_{VV}(t)}\int_{-1}^1 dy H_{I}(y,0,t).
\end{align}
Namely, the dressing of the bare vertex introduces an additional term proportional to $\delta(x)$, leading back to the modified GPD in Eq.~(\ref{eq:Hmod}). Used functions in Eq.~(\ref{eq:Hmod}) are
\begin{align}
\tilde{F}_\rho(t)&=-\frac{2 G_\rho \Pi_{VV}(t)}{1+2 G_\rho \Pi_{VV}(t)}\\
\Pi_{VV}(t)&=-\frac{N_c t}{\pi^2}\int_0^1 dy y(1-y)\mathcal{E}_1(2y-1,s)\\
\mathcal{E}_1(y,t)&=E_1\left(\frac{4M^2-t(1-y^2)}{4\Lambda_{UV}^2}\right)\nonumber\\
&\hspace{20mm}-E_1\left(\frac{4M^2-t(1-y^2)}{4\Lambda_{IR}^2}\right)\\
E_1(z)&=\int_1^\infty 
\frac{\textrm{e}^{-zt}}{t} dt .
\end{align}
Here the proper time regularization is used, with parameters determined by hadron mass spectrum and decay constant from Table 1 in~Ref.~\cite{Cloet:2014rja}, i.e., $\Lambda_{IR}=0.24$ GeV, $\Lambda_{UV}=0.645$ GeV, $M=0.4$ GeV and $G_{\rho}=11.0$ GeV$^{-2}$.

Finally, we note that the above dressing diagram doesn't modify the unpolarized TMD within the NJL model. The easiest way to see this is by realizing that in the NJL model the unpolarized PDF is obtained by integrating out the transverse momentum of TMD, i.e., $f(x)=\int d\vect{k}_T^2 f_{1}(x,\vect{k}_T^2)$. Since $f(x)=H(x,0,0)$ receives no contribution from the dressing diagrams, and $f_{1}(x,\vect{k}_T^2)$ is always positive, one deduces any corrections to $f_{1}(x,\vect{k}_T^2)$ are zero.

\bibliography{PiK3D}

\end{document}